\documentclass[preprint]{aastex}
\input{psfig.sty}

\newcommand{\degree}{$^{\circ}$}
\newcommand{\flux}{erg~cm$^{-2}$~s$^{-1}$}

\slugcomment{Accepted for publication in the Astrophysical Journal}

\shorttitle{Tanihata et al.}
\shortauthors{Long look X--ray observations of TeV blazars}

\begin{document}

%% LaTeX will automatically break titles if they run longer than
%% one line. However, you may use \\ to force a line break if
%% you desire.

\title{Variability Time Scales of TeV Blazars 
Observed in the ASCA Continuous Long-Look X--ray Monitoring}

%% Use \author, \affil, and the \and command to format
%% author and affiliation information.
%% Note that \email has replaced the old \authoremail command
%% from AASTeX v4.0. You can use \email to mark an email address
%% anywhere in the paper, not just in the front matter.
%% As in the title, you can use \\ to force line breaks.
%\author{C. D. Biemesderfer\altaffilmark{4,5}}
%\affil{National Optical Astronomy Observatories, Tucson, AZ 85719}
%\email{aastex-help@aas.org}

\author{Chiharu Tanihata\altaffilmark{1,2}, 
C. Megan Urry\altaffilmark{3}, 
Tadayuki Takahashi\altaffilmark{1,2}, 
Jun Kataoka\altaffilmark{4}, \\
Stefan J. Wagner\altaffilmark{5},
Greg M. Madejski\altaffilmark{6},
Makoto Tashiro\altaffilmark{7},
and Manabu Kouda\altaffilmark{1,2}
}

%% Notice that each of these authors has alternate affiliations, which
%% are identified by the \altaffilmark after each name.  Specify alternate
%% affiliation information with \altaffiltext, with one command per each
%% affiliation.

\altaffiltext{1}{Institute of Space and Astronautical Science,
        3-1-1 Yoshinodai, Sagamihara, 229-8510, Japan}
\altaffiltext{2}{Department of Physics, University of Tokyo,
        7-3-1 Hongo, Bunkyo-ku, Tokyo, 113-0033, Japan}
\altaffiltext{3}{Space Telescope Science Institute, Baltimore, MD, 21218, USA}
\altaffiltext{4}{Department of Physics, Tokyo Institute of Technology,
	Tokyo, 152-8551, Japan}
\altaffiltext{5}{Landessternwarte Heidelberg, 
	Konigstuhl, D-69117, Heidelberg, Germany}
\altaffiltext{6}{Stanford Linear Accelerator Center, 
	Stanford, CA, 943099-4349, USA}
\altaffiltext{7}{Department of Physics, Saitama University,
	Urawa, Saitama, 338-8570, Japan}

%% Mark off your abstract in the ``abstract'' environment. In the manuscript
%% style, abstract will output a Received/Accepted line after the
%% title and affiliation information. No date will appear since the author
%% does not have this information. The dates will be filled in by the
%% editorial office after submission.

\begin{abstract}
Three uninterrupted, long (lasting respectively 7, 10, and 10 
days) ASCA observations of the well-studied TeV-bright blazars 
Mrk~421, Mrk~501 and PKS~2155--304 all show continuous strong X--ray flaring.
Despite the relatively faint intensity states in 
2 of the 3 sources,  there was no identifiable 
quiescent period in any of the observations.  
Structure function analysis shows that all 
blazars have a characteristic time scale of $\sim$ a day, 
comparable to the recurrence time 
and to the time scale of the stronger flares.
On the other hand, examination of these flares in more detail 
reveals that each of the strong flares is 
not a smooth increase and decrease, but exhibits
substructures of shorter flares having time scales of $\sim$10 ks.
We verify via simulations that in order to explain the observed 
structure function, these shorter flares (``shots'') are 
unlikely to be fully random, but in some way are correlated with each other.  
The energy dependent cross-correlation analysis shows that
inter-band lags are not universal in TeV blazars.  
This is important since in the past, only positive detections of lags were 
reported.  In this work, we determine that the sign of a lag may 
differ from flare to flare;  
significant lags of both signs were detected from 
several flares, while no significant lag was detected from others.
However, we also argue that the nature of the underlying component 
can affect these values.  
The facts that all flares are nearly symmetric
and that fast variability shorter than the characteristic
time scale is strongly suppressed,
support the scenario where the light crossing time dominates the 
variability time scales of the day-scale flares.  

\end{abstract}

%% Keywords should appear after the \end{abstract} command. The uncommented
%% example has been keyed in ApJ style. See the instructions to authors
%% for the journal to which you are submitting your paper to determine
%% what keyword punctuation is appropriate.


%% From the front matter, we move on to the body of the paper.
%% In the first two sections, notice the use of the natbib \citep
%% and \citet commands to identify citations.  The citations are
%% tied to the reference list via symbolic KEYs. The KEY corresponds
%% to the KEY in the \bibitem in the reference list below. We have
%% chosen the first three characters of the first author's name plus
%% the last two numeral of the year of publication as our KEY for
%% each reference.

\section{INTRODUCTION}

Jets are among the most exciting (but also among the least understood)
cosmic phenomena, being unusually efficient particle accelerators
that can generate relativistic electron distributions, 
which in turn produce synchrotron and inverse Compton radiation.
The jet power is generated near the central black hole,
probably via the conversion of gravitational energy from accreting matter
by an as-yet unknown mechanism.  Blazars are Active Galactic 
Nuclei (AGN) possessing jets aligned close to
the line of sight, and thus the observed emission is dominated by the
Doppler-enhanced jet component 
(e.g. Blandford \& Konigl 1979; Urry \& Padovani 1995).
This makes blazars ideal targets for studying jet physics.  

Rapid variability or flaring, which is one of the major 
characteristic of blazars, provides important clues towards 
understanding the conditions in the jet.  Variability of the 
synchrotron spectral component is most rapid above the 
$\nu F_\nu$ peak of the emission \citep{ulrich97}, which is 
measured in the infrared to X--ray energy range.  X--rays provide 
an ideal means for studying variability because there is little
contamination from emission sources other than the jet 
(e.g., galaxy or nuclear emission), and because integration times
are limited only by statistics. 

The blazars with peak synchrotron output in the X--ray range also emit
strongly at $\gamma$--ray energies.  The brightest of those have been 
detected in the TeV range with ground-based air-shower arrays. 
These so-called ``TeV blazars'' include Mrk~421 \citep{punch92} 
and Mrk~501 \citep{quinn96} as the brightest sources, 
and several other objects including PKS~2155--304
\citep{chadwick99} at somewhat fainter TeV fluxes.        
In the TeV blazars, the X--ray emission probes the electrons 
accelerated to the highest energies;  those electrons 
have the shortest synchrotron cooling times.
The TeV blazars listed above have been well studied previously
at the X--ray energies via numerous multi-wavelength monitoring campaigns
(e.g., Mrk~421: Macomb et al.\ 1995, Maraschi et al.\ 1999,
Takahashi et al.\ 2000; Mrk~501: Catanese et al.\ 1997,
Kataoka et al.\ 1999, Sambruna et al.\ 2000;
PKS~2155--304: Urry et al.\ 1997)
However, it was only the recent $ASCA$ ``long look'' 
at Mrk~421 \citep{tad00}, a 7-day uninterrupted continuous 
observation in April 1998, that showed rapid daily flares not 
previously resolved with sparser sampling.  Indeed, the wealth of 
information from this $ASCA$ observation strongly underlined the 
need to obtain similar high quality data for other blazars. 

To optimize $ASCA$ science in its eighth year, especially as the 
new generation of X--ray telescopes was becoming available, the AO-8 
program was explicitly devoted to long observations of interesting 
targets. We therefore proposed two 10-day campaigns for the next two
X--ray brightest TeV blazars, Mrk~501 and PKS~2155--304.
Our goal was to probe the physical conditions in the blazar jets.  
When events in the jet (such as shocks or changes in the magnetic field)
cause changes in the electron energy distribution, we see
corresponding changes (such as flares) in the emitted X--rays. From 
the energy dependence of these variations, we can infer
constraints on the time scales for acceleration, injection, and cooling,
and on the size of the synchrotron-emitting region.
Adding two more blazars to the data already obtained for Mrk~421,
we have a well-studied sample of three TeV blazars, from which 
we can begin to study the range of properties of blazar jets.

We describe the $ASCA$ observations and data reduction in \S~2.
In \S~3, we discuss 
the analysis of the time series via the structure function method;  
this includes simulation of light curves as consisting of series of 
``shots.''  We compare the structure functions calculated for 
the observed and simulated data in order to verify what are the 
preferred parameters of such ``shots.'' We discuss the variability 
time scales in \S~4, address the question of the energy 
dependence of temporal variability through cross-correlations in \S~5,
and present our conclusions in \S~6.  

\section{OBSERVATIONS AND DATA REDUCTION}
We observed Mrk~421 for 7 days with $ASCA$ in 1998 
during April 23.97 -- 30.8 UT, 
and for 10 days each of Mrk~501 and PKS~2155--304 during
2000 Mar 1.50 -- 11.00 UT and  2000 May 1.55 -- 11.50 UT, respectively.
For all observations, the SIS (Solid-state Imaging Spectrometer; 
\citet{burke91,yamashita97}) 
was operated in 1~CCD FAINT mode for the high data rate, 
and BRIGHT mode for the medium/low rate. In order to combine with
the BRIGHT mode data and apply standard data analysis procedures,
the FAINT mode data were converted to BRIGHT mode data.
Similarly, the GIS (Gas Imaging Spectrometer; \citet{ohashi96})
was always operated in the PH-nominal mode.

All data reduction was performed using the HEASOFT 5.0 software package.
The screening criteria for both the SIS and GIS included rejection of 
data during passage through the South Atlantic Anomaly and 
for geomagnetic cutoff rigidity lower than 6 GeV/$c$.
For the SIS, we used regions with angle from
bright Earth greater than 20\degree, and 
angle from night Earth larger than 10\degree.  We also selected 
only the X--ray events according to how the
charge was distributed to the CCDs, namely 
these corresponding to grades 0, 2, 3, and 4.
For the GIS, we accumulated data for bright Earth and night Earth elevation
angles larger than 5\degree.

Source photons were extracted from circular regions centered at 
the target, with 3$'$ and 6$'$ radii for SIS and GIS, respectively
for Mrk~501 and PKS~2155--304.  As described in \citet{tad00}, 
we use only the SIS for Mrk~421 since the source was sufficiently 
bright that there was telemetry saturation in the GIS.  
We used photons only from the smaller regions, corresponding to  
1$'$ and 2$'$.6 radii for SIS0 and SIS1, respectively,
in order to avoid saturation in the SIS cameras.
The background data were examined using source-free 
regions in the same image. Because the background count rates 
and their fluctuations were negligible (less than 2 \% at most), 
we did not perform any background subtraction before
further analysis, so as to avoid introducing any artifacts.
For the X--ray light curves, we combined as much data as feasible 
(i.e. 2 SISs for Mrk~421 and all SIS and GIS for the other two sources)
to maximize the signal-to-noise ratio in each temporal bin.

\subsection{Mrk~501}
When the 10-day observation of Mrk~501 started, the object was
in a moderately faint state, with its 2 -- 10~keV flux at 
$\sim 6.0 \times 10^{-11}$~\flux, compared to an historical
range of $\sim (1 - 60) \times 10^{-11}$~\flux \citep{dellaceca90,pian98}.
The spectra integrated over short time periods were 
well fitted by power laws with photon index $\Gamma$ = 1.8 -- 2.2,
well within the range previously observed.  
The spectrum indices suggests that the synchrotron peak is
shifting over a wide range, from below to above the 
$ASCA$ bandpass, during our observation.
%The relatively flat spectrum suggests the synchrotron peak is 
%near the $ASCA$ bandpass.

The 10-day light curve is shown in Figure~\ref{fig:lc_mrk501},
where the heavy points represent the soft X--ray band 
(0.6 -- 2~keV), and the light  points represent the
hard X--ray band (2 -- 10~keV) in 5678 sec bins,
equal to the satellite orbit period during the observation. 
Each light curve was normalized to its mean.
After 2 days of small variations, there was a large flare,
followed by 2 additional well-defined 
flares, and a further rise at the end of the observation.  
Even between the large flares there is additional, significant
variability, so that Mrk~501 is never in a truly quiescent state.
The amplitude of variability in the hard X--ray band is clearly larger,
indicating a hardening of the spectrum during high states.
The flares in the two bands are well correlated.  In particular, 
near the maximum of the flare, there are 
apparently no lags larger than a few temporal bins 
(i.e., lags $\lesssim 10^4$~s).

Examining each flare in more detail, 
we observed that none of the large flares show a smooth rise or
decay but exhibit substructures, with smaller flares 
having shorter time scales.
This is demonstrated in the small panel which shows a
blow-up of the last flare (marked with an arrow), 
where the hard energy band light curve is binned to 2048 sec.
The overlapping, or closely placed flares are clearly resolved,
and the rise and decay time scales are of the order of 5 -- 10 ks.  
The amplitude of the substructures are equal to or larger than 10 \%.
We remark that in $ASCA$, and in particular in our
case where the background level is
less than 2 \% throughout the observation,
the statistical error completely dominates the systematical error,
and thus the substructures cannot be due to background fluctuation etc.

All the large flares have different durations
ranging from 40 to 100 ks, but importantly 
all flares have nearly symmetric time profiles, 
similar to those seen in the long-look observation of Mrk~421.
The new result is that the substructures, lasting on 
the order of 5 -- 10 ks, also all appear to be rather symmetric;
no faster rise or decay than that time scale is apparent.
However, the complicated superposition of flares makes it hard 
to make a quantitative statement regarding the symmetry of each flare.
Further analysis regarding the symmetry using higher order moments will 
be presented in a separate paper.

\subsection{PKS~2155--304}
The observation of PKS~2155--304 also started with the source 
in a relatively faint state, with its 2 -- 10~keV flux 
at $\sim 3.0 \times 10^{-11}$~\flux,
compared to the historical range spanning 
$\sim (1.5 - 25) \times 10^{-11}$~\flux 
\citep{dellaceca90,kataoka00}.
The time-resolved spectra, integrated over short time periods, were 
again well fitted by power laws with photon indices ranging 
$\Gamma$ = 2.5 -- 2.8, consistent with previous observations.
Here the synchrotron peak must be below $\sim 1$~keV, and the
observed X--rays represent the high-energy tail of the electron 
energy distribution.  

The $ASCA$ light curve of PKS~2155--304 is shown in 
Figure~\ref{fig:lc_pks2155}.
Again, the heavy points represent the soft X--ray band 
(0.6 -- 2~keV), and the light points represent the
hard X--ray band (2 -- 10~keV), in 5657 sec bins.
Each light curve was normalized to its mean.
In this case, the intensity increased strongly towards the end
of the observation, by nearly a factor of 3. 
This high amplitude is consistent with the steepness of the spectrum
since we expect stronger variability above the synchrotron peak, 
where radiative losses dominate (if the peak represents an
equilibrium between loss and acceleration processes).

In addition, there is continuous flaring throughout the 10-day 
observation, each having amplitude $\sim30$ -- 50 \%.
Here the variability amplitudes in the hard and soft X--ray bands
are very similar, indicating only a modest spectral change even during 
the large amplitude flare. The variability in the two energy bands 
is well correlated, and again concerning the flare maximum,
any lags between the two bands must be smaller than $\sim 10^4$~sec.

A close look at the light curve again shows substructures
present in the large flares.  This can be seen in the small panel
which shows a blow-up of the last part of the flare, using
a 2048 sec binned  light curve.  
Shorter variability with rise and decay time scales of $\sim 10$ ks 
is clearly seen.

\subsection{Mrk~421}
The full 7-day light curve of Mrk~421 is shown in 
Figure~\ref{fig:lc_mrk421}.
The heavy points represent the soft X--ray band 
(0.6 -- 2~keV), and the light points represent the
hard X--ray band (2 -- 10~keV), in 5740 sec bins.
Each light curve was normalized to its mean.
This was the first TeV blazar where repetitive daily flares were observed.  
Details of the daily flares in Mrk~421 are presented in \citet{tad00}.  

The remarkable brightness of the source during the
observation enabled us to 
further study the structure of the light curve in 512 sec bins.
And for this source as well, 
the large daily flares clearly show substructures,
as shown in the smaller panel.
The time scale seems to be even shorter in Mrk~421 than in the other 
two objects, and the shortest flares that can be detected 
show significant variations with time scales of $\sim 5$ ks.  

\section{STRUCTURE FUNCTION ANALYSIS}
\subsection{Structure Functions Derived from Observations}
The continuous coverage and long duration of the $ASCA$ observations
provide the best opportunity to date for studying
variability in blazars, enabling us to track the individual flares as 
well as the underlying component, and to determine the rise or decay
time scales.  However problems still remain since it is 
always difficult to unambiguously determine from the light curve 
whether a given observed flare is a single shot, or if it consists 
of shorter overlapping shots.

Another advantage of continuous long observations is that 
it is possible to employ the structure function, which can provide a
quantitative statistical test of the time series.  We use the formalism 
described in \citet{simonetti85};  one modification 
 is that since photon statistics dominate the error 
in X--ray data (i.e. error depends on flux, unlike low frequency 
observations where systematic error dominates), we use a continuous 
weighing factor proportional to its significance of each data point
instead of taking only either 0 or 1 as in \citet{simonetti85}.
The definition of the 1st order structure function for a 
light curve described as $f(i)$ is then,
\begin{equation}
D_f^1(k) = \frac{1}{N^1(k)} \sum w(i)w(i+k)[f(i+k)-f(i)]^2 ,
\end{equation}
where
\begin{eqnarray}
N^1(k) &=& \sum w(i)w(i+k) , \\
w(i) &\propto& \frac{f(i)}{\sigma_f(i)} .
\end{eqnarray}
Here, $w(i)$ is the weighing factor, and 
$\sigma_f(i)$ is the 1 $\sigma$ uncertainty of the
data point $f(i)$ in the light curve.
The summations are made over all pairs.

We set the light curve bin-size equal to the
orbital period as used in Figure 1--3,
so as to minimize the effect of short gaps 
mainly due to source occultation and passage through 
the SAA or high particle background regions.
The orbital period was 5740.5 sec for Mrk~421, 
5678.3 sec for Mrk~501, and 5657.2 sec for PKS~2155--304. 
The decrease of the orbital period in 10 days is short, no more than 
$\sim$ 5 s, and thus it is not taken into account.
The calculated structure functions for the three sources are 
shown in Figure~\ref{fig:sf}(a)--(c).  We must remark that the 
``wiggling'' feature at the longest time scales
are due to the finite number of flares that exist in the
observed light curve, and thus are not intrinsic to the source.
The number of pairs in the summations in equation (1) decreases 
as the time difference increases, and accordingly the uncertainty 
becomes larger (see Kataoka et al.\ 2001 for a detailed discussion).

The first remark is the steep slope at the short time scales in all 
sources.  The slope $\beta$ of a structure function is believed to be 
an indicator of the nature of the intrinsic variation of the source;
e.g. $\beta=1$ for red noise (random walk), $\beta=0$ for white noise.
The slope at the shortest end of the separation time, 
$\beta\sim$1.5 for Mrk~501, $\beta\sim$1.4 for PKS~2155--304, 
and $\beta\sim$1.4 for Mrk~421, are all
steeper even than the ``random walk'' case, meaning there is a
rapid decrease in variability amplitude with temporal frequency.  
Somehow the variable source retains a ``memory'' of its state
over a certain period.  Another remark is the flattening of the 
slope at longer time scales, indicating little increase in variability 
amplitude.  The observed breaks (characteristic time scale)
are at $\sim1$~day for Mrk~501, $\sim 2-3$~days for PKS~2155--304,
and $\sim0.5$~day for Mrk~421 
(the second would be determined more precisely with a longer observation).

\subsection{Simulated Light Curves and Structure Functions}
The structure function analysis suggests that for all TeV blazars 
these functions have common features, a steep slope on the short time scales 
which flattens at longer time scales.  On the other hand, it is 
not obvious what exactly these slopes and breaks indicate.
The likely time scales to contribute would be the rise and decay 
time scales of each flare, and also the separation of the flares.
Here we simulate various light curves and calculate the structure 
function in order to determine  which parameters are most likely to 
affect the slopes and break time scales measured in the structure 
functions derived from the data.  

In simulating the light curves, we regard the whole light curve as 
a superposition of various flares or ``shots'' occurring randomly, following 
a Poisson distribution.  (An offset component is also possible, 
but this has no effect on the structure function.)  The shape of 
each shots is difficult to determine from observation, 
but here we assume an simple triangle shape, where $\tau_{\rm r}$ is the 
rise time scale, $\tau_{\rm d}$ is the decay time scale, 
and $t_{\rm p}$ is the time of occurrence of the shot.  With this, 
\begin{eqnarray}
f(t) = \left\{
\begin{array}{ll}
	A \, (t-(t_{\rm p}-\tau_{\rm r}))   & {\rm for}\; t \leq t_{\rm p} \\
	-A \,(t-(t_{\rm p}+\tau_{\rm d}))   & {\rm for}\; t > t_{\rm p} .
\end{array}
\right.
\end{eqnarray}
We note that the results obtained in the following sections 
concerning the slope and the location of the break time scale 
did not change if we instead assumed a Gaussian or an exponential 
time profile.

\subsubsection{Dependence on $\tau$}

First, we consider the case where all shots have the same 
time scales $\tau_{\rm r} = \tau_{\rm d} = \tau$.
We set $\tau$=10 and 100, 
and simulated 2000 steps as the light curve, respectively.
The shots occur randomly, with the 
average number of shots being 1 per time unit.
The intensity of each shot is also set randomly, 
but the number density follows a power law distribution
N $\propto$ $A^{-1.5}$, where $A$ is the normalization from
equation (4).  This distribution was chosen to be 
similar to gamma--ray bursts \citep{pendelton96},
but this again is not sensitive to the slope 
and break time scale of the resulting structure function.
This will be used commonly throughout this section.

A set of simulated light curves for $\tau = 10$ and
$\tau = 100$ are shown in Figure \ref{fig:sim-lc} (a-1) (a-2),
and the calculated structure functions are shown in 
the top panel of Figure \ref{fig:sim-sf}(a).
The shape is similar to that calculated from the observations;  
a steep index at the short time scales and a flattening at longer time scales.
The location of the break is clearly seen to shift together 
with the time scale of the individual shots in each light curve.
The bottom panel shows the power-law index of the structure 
function, where we can see that the index of the steep slope 
reaches $\beta$=2 at the shortest time scale, and continuously 
flattens to $\beta$=0 towards the time scale $\tau$.

\subsubsection{Dependence on Rate} 

In the second case we fixed $\tau = 50$, and 
varied the average number of shots per time unit $R = 0.1$, 1, and 10.
It is often the case that the variability amplitude 
of blazar flares is of the same order as the offset-like underlying
component. It is not yet understood whether this is
due to a flaring component superposed on a steady emission component 
(which requires continuous acceleration), 
or many flares generating the offset component.
In fact, as can be seen in the simulated light curves for
each case shown in Figure \ref{fig:sim-lc} (b-1) (b-2) (b-3),
when the rate is set high, the visible flare amplitude becomes 
comparable to the underlying offset.

The resulting structure function and the index of the slope 
is shown in Figure \ref{fig:sim-sf}(b).  Here, only the normalization 
is changed, and the break remains at the same time scale not 
depending on the shot cycle.  The steep slope reaching $\beta$=2 
is also common.

\subsubsection{Simulations with Variable $\tau$}
Now we consider a range of $\tau$ for individual shots.
Figure \ref{fig:sim-sf}(c) is the resulting structure
function when $\tau$ takes a random value 
between $\tau_{\rm min} = 10$ and $\tau_{\rm max} = 100$.
The rate is set to $R = 1$.  The structure function of the 
case with a  single $\tau = 100$ is also plotted for comparison.
The steep slope $\beta$ $\sim$ 2 at the shortest time scale
is similar to the case of a single $\tau$, but
it becomes flatter after $\tau_{\rm min}$,
and becomes completely flat after  $\tau_{\rm max}$.
The slope between  $\tau_{\rm min}$ and  $\tau_{\rm max}$ is $\beta$ $\sim$ 1, 
The simulated light curve is shown in Figure \ref{fig:sim-lc}(c-1),

\subsubsection{Non-symmetric Shots}

The last case considered by us is when the rise and decay of the 
individual shots are different.  We fixed the decay time scale 
$\tau_{\rm d}$=100 and varied the rise time scale 
$\tau_{\rm r}$ = 1, 10, and 100.  The rate is set to $R = 1$.
The simulated light curves are shown in Figure \ref{fig:sim-lc}(d-1)(d-2),
and Figure \ref{fig:sim-sf}(d) is the resulting structure function.
The time scale where the structure function completely flattens 
to $\beta = 0$ is common in all cases, but the
slope in shorter time scales appears to vary.
This is, however, exactly analogous to the above case
where $\tau_{\rm r} = \tau_{\rm d} = \tau$ had a range.
The index becomes 2 below the shortest $\tau$ and 
rises with $\beta$ $\sim$ 1 up to the longest $\tau$.
Since the observed light curve 
imposes a limit on the detectable shortest time scale, 
the observed slope in the structure function is consistent with a range 
of $\beta$.  Accordingly, the steepest slope observed appears 
to be determined by the relation of the minimum observable time scale and 
the shorter time scale of the rise or decay.

\subsection{Summary of the Simulations and the Detection Limit with 
the Structure Function}

The results derived in the simulations above can be 
summarized as below. We recall that the assumption is that
the whole variation (besides a DC component) is a result of
randomly occurring ``shots'' superposed on each other.
\begin{enumerate}
\item The time scale of the individual shots $\tau$
determines the location of the break in the structure function,
where the power-law slope approaches $\beta$ $\sim$ 2 towards the 
shorter time scale.
\item Result (1) doesn't depend on the shot occuring rates.
\item When $\tau$ takes a range of time scales, the break (from index of 2)
occurs at the minimum time scale of those shots.
\item When the rise and decay time scales of the shots are different,
the break occurs at the time scale that is shorter of the two.  
\end{enumerate}
The above considerations imply that 
when there is a minimum time scale $\tau$ for individual shots,
the slope of the structure function becomes steep below 
$\tau$, and approaches $\beta$ $\sim$ 2.
For the longer time scales, the structure function
increases up to where there is no more
shots having that time scale.

On the other hand, it is also true that the structure functions 
are by definition being dominated by shots with larger amplitude.
We tested this with adding very small amplitude and
shorter time scale shots having $\tau = 10$ on top of a 
light curve with a single $\tau = 100$ as discussed above.  
A portion of the light curve is shown in 
Figure \ref{fig:sim-sf2}(a) and (b).
Although the structure functions consisting of shots with 
individual values of $\tau$ would 
depend on $\tau$ as in Figure \ref{fig:sim-sf}(a), 
the structure function of the summed light curve 
is very close to the one with only the larger long shots.
Thus if there is a situation where there are numbers 
of very short shots having very small amplitude,
together with a longer and larger amplitude variation,
such small, rapid shots will not affect the structure function.

The situation would be the same when the time scale of the shots
decreases with amplitude.   Such a case can be envisioned 
if the electron density is equal for each emission region 
in the jet:  the region size would then be smaller for smaller shots.
Thus the important result is that it is that we cannot reject 
the existence of a variable component having shorter time scales than
the break time scale, but such a component, which would be characterized 
by short time scale, cannot dominate the variability in the light curve.

\section{DISCUSSION OF THE TIME SCALES}

The long look observations of TeV blazars showed that daily 
flares are common to all sources, and that 
there was actually no particular quiescent period throughout 
our observations.  The observation of Mrk~501 and PKS~2155--304 
in 2000 caught the sources in relatively faint states compared 
to the historical values of their flux, which also demonstrated that a 
high state is not a requirement for rapid variability.  Another 
important result is that each of the day-scale flares is actually 
not a single, isolated flare, but shows substructures with shorter 
time scales of $\sim 5$ -- 10 ks. This was observed commonly in our 
long-look observations for all sources.  We will further discuss 
this short time scale variability at the end of this section.

Structure function analysis showed that the X--ray light curves
of all TeV blazars have two common features:  a very steep slope 
at the shortest time scale (steeper than red noise), and a break 
at the time scale of $\sim$ a day, where the slope flattens toward 
longer time scales.  As shown from the simulations above, 
the location of this break (where the slope turns over from a 
power-law index of $\beta=2$), is indicative of time scales characteristic 
of individual flares.  The observed steepest power-law slope at the 
shortest time scales was $\beta = 1.5$, 1.4, and 1.4 for Mrk~501,
PKS~2155--304, and Mrk~421, respectively.  This slope is likely 
to reach the value of $\beta=2$ at even shorter temporal frequencies.
In fact, we derived a steeper slope at even shorter time scales 
by using a light curve in shorter time bins, $\beta \sim1.8$ for Mrk~421;  
probing these shorter time scales in Mrk~501 and PKS~2155--304 becomes 
more difficult since the photon counts are smaller, and the effect of 
the measurement error cannot be completely removed.

Estimating the exact location of the break requires
careful simulations as described in \citet{kataoka01},
but the important result here is the existence of 
a common break in the structure functions.
This was suggested in \citet{kataoka01} by collecting 
all previous $ASCA$ and $RXTE$ observations up to 1999.
However, an actual confirmation of this suggestion required 
a single, continuous, long-look observation in the X--ray 
range, which we presented above.  

Another observational result from the $ASCA$ long looks at 
TeV blazars in the X--ray band is that the flares are always 
nearly symmetric.  This is often interpreted by the 
light-crossing time dominating the variability time scale
and thus smearing out the other time scales \citep{chiaberge99}.
In our simulations where the time series consists of superposition 
of individual ``shots,'' the characteristic time scale in the
structure function is likely to reflect the time scales of the
individual shots:  with this, the light-crossing times are most 
likely to determine the characteristic time scale in the structure 
functions.  Furthermore, because the flares do not have flat-top 
profiles, this suggests that the cooling and acceleration time 
scales are comparable to the light-crossing time.  The cross-correlation 
analysis given below 
showing significant lags of both signs (soft-lags and hard-lags) 
also supports the inference that the cooling and acceleration time 
are comparable.  

The cooling time (in the observer's frame) of the electron 
emitting synchrotron photons with energy E$_{\rm keV}$ can be calculated as 
$\tau_{\rm cool} \simeq 1.7 \times 10^4 \, 
(1 + u_{\rm soft} / u_{\rm B})^{-1} \, 
\delta_{10}^{-0.5} \, B_{0.1}^{-1.5} \, E_{\rm keV}^{-0.5}$,
where $B_{0.1}$ is the magnetic field in units of 0.1 Gauss, 
and $\delta_{10}$ is the beaming factor in units of 10.
Here the  peak frequency of the synchrotron spectrum 
is given by $\nu \sim 1.2 \times 10^6 \,B \,\delta \,\gamma^2$.
(e.g., Rybicki \& Lightman 1979).  Fitting the multi-frequency spectra, 
\citet{kataokaD} derived the magnetic fields $B$ 
and beaming factors $\delta$ of respectively 
$B$ = 0.13 Gauss, $\delta$ = 14 for Mrk~421, 
$B$ = 0.13 Gauss, $\delta$ = 9 for Mrk~501, and 
$B$ = 0.14 Gauss, $\delta$ = 28 for PKS~2155--304.  With this, 
we infer the cooling times for the electrons corresponding to 
1 keV photons to be $\sim$ 5000, $\sim$ 6000, and $\sim$ 3000 s,
respectively.  
Although some flexibility still remains in these model parameters,
and even considering that all parameters may change by a factor of 2 
\citep{fab98,kataoka99,kataokaD},
the cooling time is likely to fall in the range of 1 -- 10 ks.
Also because these parameters were derived for different epochs, 
they may not be exactly identical during the flares observed in
the long look observations. However, given the
fact that the values derived for various epochs are not 
that different, it is natural to think that the parameters 
during our observations are quite similar to the values given above.
Thus, assuming that the observed characteristic time scale
(i.e. $\sim$ a day) is reflecting the light-crossing time of 
the emission region, this calculation indicates that 
the cooling time of the electrons emitting the X--ray photons is
significantly shorter than the light crossing time.

The origin of flares observed in blazars is far from being fully 
understood, but one of the simplest scenarios would be an increase
in the numbers of electrons injected into the acceleration region.  
This may well be a region where a shock is 
passing through some region where the electron density is high 
(e.g. Kirk, Rieger, \& Mastichiadis 1998).  These electrons will 
escape from the acceleration process and enter the emission region 
and cool.  In this scenario, the overall emission region
would be defined by the size of the region where the number of 
injected electrons is high.  Furthermore, if this emission region 
has a size characterized by a time scale of $\sim$ a day,
the number of injected electrons is unlikely to be constant
since the cooling time of the electrons is short.  
With this, in order to have a flare without 
a flat-top when the acceleration and cooling times 
are much shorter than the light-crossing time, it is likely that 
electron density is somewhat inhomogeneous, and/or it is an effect of 
a more complex geometry (such as a sphere, instead of a shell with 
constant thickness).

One of the important observational results discovered in our 
long-look observations is the common existence of substructures,
having shorter variability time scales, on the order of $\sim$10 ks
(hereafter referred to as '$\sim 10$ ks flares'), apparent via 
closer examination of the larger day-scale flares in the observed 
light curve. 
This was also observed in the recent 
light curve of Mrk~421 taken by the $XMM$-$Newton$ satellite 
which showed with great statistics
overlapping flares with time scales 
of $\sim$ 5 -- 10 ks \citep{brinkmann01}. 
It is not obvious whether these $\sim 10$ ks flares are 
superposed on a longer and larger amplitude flare, or if the entire 
variability is a result of a superposition of a number of these 
$\sim 10$ ks flares piled up on top of each other.  If individual 
shots are to have this time scale of $\sim 10$ ks, reflecting the 
light-crossing time, the cooling and acceleration times would be 
in good agreement with the light crossing time, which would be in turn 
consistent with the observed symmetric non-flat-top flares.

Our simulations showed that if the entire variability consists of 
shots with time scales of 10 ks occurring randomly, a break in the 
structure function (indicating the characteristic time scale) 
will be located at 10 ks (see Figure~\ref{fig:sim-sf}).
Even if such time series consisted of a mixture of shots having
randomly distributed time scales with a minimum at 10 ks,
the location of the break will still be at 10 ks.  However, as was 
shown in Figure \ref{fig:sf}, observations definitely show a 
characteristic time scale at $\sim$ a day.  This implies that 
if all shots have the time scale of 10 ks, the occurrence of such 
shots should not be fully random, but by somehow correlated with 
each other, generating the observed characteristic time scale of 
$\sim$ a day.  This can be either correlated in time, or in space.  

If shots  with longer time scale also exist, there must be a reason for 
the shorter shots having smaller amplitude than the longer ones,
since otherwise, the break time scale in the structure function 
would be located at 10 ks.  Small variations superposed on the 
larger flares having the characteristic time scale may also be a 
likely explanation.  Our $ASCA$ observations do not have enough 
statistics to fully resolve and study the energy dependent variability
in the smaller flares.  Continuous monitoring with telescopes with 
larger photon collecting area, such as $XMM$-$Newton$, would surely 
help further understandings of the shorter time scale variability.

\section{CROSS CORRELATION AND LAGS AS A FUNCTION OF ENERGY}

The goal of X--ray monitoring of blazars is to understand the 
underlying jet physics. The time scales of variability and
spectral changes across flares can be interpreted in terms of
the time scales for electron injection, acceleration, and cooling,
taking into account propagation of light across the emitting volume
and from the source to us.
For example, soft lags, i.e. the hard energy band leading the
variation, have been interpreted as signatures of
synchrotron cooling \citep{tashiro92,tad96,urry97,kataoka00}.
Intensive X--ray monitoring of blazars
has revealed not only soft lags but in some cases hard lags
\citep{sembay93,tad00,catanese.rita00}, which
may be interpreted in terms of the energy dependence in
acceleration time scales.  The physical situation is likely 
more complex than assumed in the simple cooling scenario.

To explore further the energy dependence in the variability 
than is readily apparent from the light curve in Figure 1--2,
we make use of the discrete correlation function (DCF) method 
developed by \citet{edelson.krolik88} and also 
the modified mean deviation (MMD) method 
originally introduced by \citet{hufnagel.bregman92}.
The advantage of the DCF compared to the classical correlation 
function is that it is applicable to unevenly sampled data,
thus only the existing data are used, requiring no interpolation.
For the calculations we binned the light curve in bin sizes of 1024 s.

We must remark that ``lags'' studied in this paper 
do not necessarily indicate one energy band being 
a delayed version observed in the other band.
This is rather different from situations of comparing 2 time series 
where one is thought to be driven by the other, such as 
the lines and continuum (e.g. Edelson \& Krolik 1999;
Vaughan \& Edelson 2001) or thermal reprocessing 
(e.g. Edelson et al.\ 1996; Nandra et al.\ 2000) in Seyfert galaxies. 
The time series we are comparing here are certainly due to the 
same electron population cooling via a common radiation process; 
in our case, this is primarily synchrotron emission, where the rate of 
electron energy loss is dependent on that energy.
(Compton cooling for this energy range of electron
is strongly suppressed by the Klein-Nishina cutoff.
e.g. Kataoka 2000; Li \& Kusunose 2000)
If cooling or acceleration is the cause of lags as it is often 
interpreted, the difference is likely to arise in the decay 
or rise time scales of the flares, and what we want to verify 
and quantify is the energy dependence.  In this case, the DCF 
or MMD is likely to be asymmetric indicating the differences of 
the time scales, and the deviation from zero lag is still useful to
parameterize the energy dependence of the time scales.  After 
investigating the cross correlations, we comment on the importance 
of evaluating the significances of the values of the lags.

\subsection{Analysis of Complete Time Series}

We first considered the full light curve data sets.
We separated each light curve into 4 energy bands,
0.6 -- 1.2 keV, 1.2 -- 1.8 keV, 1.8 -- 3 keV, and 3 -- 10 keV, 
and calculated the DCF/MMD of the lower three energy bands
against the 3 -- 10 keV band.
The case of comparing of the 0.6 -- 1.2 keV and 3 -- 10 keV data 
is shown in the middle and bottom panels in Figure \ref{fig:ccf_full}
for all three sources.  We can see that there is strong correlation 
with $r_{\rm max} \geq 0.8$ for all sources. Here,
$r_{\rm max}$ is defined by the maximum of the correlation peak.
In contrast to the common strong correlation,
the width of the correlation peak appears different in each source, 
with PKS~2155--304 being the widest of the three.  
This comes from the fact that cross-correlation functions
are convolutions of autocorrelation functions with non-zero width. 
This can be seen by comparing with the autocorrelation
functions (ACFs) measured using the DCF technique shown in the 
top panels of Figure \ref{fig:ccf_full}.
The width of the ACF peak is smallest in Mrk~421 ($\sigma \sim$ 15 ks,
when fitted with a Gaussian function),
second in Mrk~501 ($\sigma \sim$ 30 ks), and
largest in PKS~2155--304 ($\sigma \sim$ 60 ks). 
In the previous sections, we argue that 
all TeV blazars have a characteristic time scale
which is likely reflecting the time scales of the day-scale flares.
It is interesting to note that 
the characteristic time scales 
derived in the structure function analysis in section 3,
are commonly 2--3 times of the width (defined as the $\sigma$
as above) of the ACF peak.

We estimated the lag by fitting a Gaussian plus constant model
(see discussions in Edelson et al.\ 1995; Peterson et al.\ 1998)
to each DCF and MMD.  The results of the energy dependent lag 
for the three sources are presented in Table~\ref{tbl:lag_full}.
The positive value indicates the soft band lagging the hard band.
Here, PKS~2155--304 was the only source to show a
soft lag with the ``expected'' energy dependence, 
increasing value with larger energy gap.
There was no evidence of lags using the 
entire light curve in the Mrk~421 or Mrk~501 data 
(consistent with no lag within the statistics).
The errors represent the 1 $\sigma$ uncertainty
estimated by realistic Monte-Carlo simulations, using mainly
the flux randomization (FR) and random subset selection (RSS)
methods described in \citet{peterson98}.
All uncertainties concerning lags will be estimated in the
same way throughout the paper.

\subsection{Analysis of Individual Flares}

Following the examination of the entire time series, we also 
studied individual flares.  Here we restrict our discussion 
to the flares with the rise and decay times on the order of the 
characteristic time scale of the source derived from the structure 
function analysis, thus the time scale dominating the variability power
(i.e. the day-scale flares, not the small hour-scale substructures).
The results for Mrk~421 during the long-look campaign are described 
in \citet{tad00}, and thus here we only discuss in detail the 
results for Mrk~501 and PKS~2155--304.  For Mrk~501 this was 
straightforward, since the light curve can be divided clearly into 
segments spanning each of the first three large flares, and the 
rise part at the end.  Regarding the time regions we took the data 
corresponding to the running time of the observation of 
160 -- 340 ks, 440 -- 520 ks, and 570 -- 600 ks, and we refer to those 
respectively as flare1, flare2, and flare3.  The calculated DCF 
and MMD between the 0.6 -- 1.2 keV and 3 -- 10 keV band for each 
flare are shown in Figure \ref{fig:ccf_mrk501}.  Again, the lag 
was first estimated by fitting a Gaussian plus constant model.  
This is shown as the solid line in Figure 
\ref{fig:ccf_mrk501}.  The results of the energy dependent lag 
for the three flares are summarized in Table~\ref{tbl:lag_flare}(a).
Interestingly, in contrast to the full light curve of Mrk~501 
which showed no detectable lag, the individual flares had different 
features.  Flare1 showed a soft lag, and flare2 a hard lag, although 
the significance is small for flare1 (2.5 $\sigma$ for the 
lowest and highest energy band).  Flare3 also shows a trend 
of a soft lag, but the significance is also small.
The values of the lags are in the range of $\sim 1 - 10$ ks,
which is $\lesssim 2$ bins of the counts light curves plotted in 
Figure~\ref{fig:lc_mrk501}.

The flares in PKS~2155--304 are more complicated (with overlapping 
smaller flares), but concentrating on the flares dominating the 
variability power, we divided the light curve into 3 time regions,
0 -- 200 ks, 200 -- 500 ks, and 500 -- 840 ks (referred to as flare1, 
flare2, and flare3, respectively).  The calculated DCF and MMD 
for the 1.2 -- 1.8 keV and 3 -- 10 keV bands for each flare are 
shown in Figure \ref{fig:ccf_pks2155}.  Similarly, lags were estimated 
by fitting a Gaussian plus constant model as shown together as solid 
lines.  The results of the energy dependent lag for the three flares 
are summarized in Table~\ref{tbl:lag_flare}(c).  Here, flare1 and flare2 
showed a energy dependent soft-lag, where no detectable lag was seen in 
flare3.

In several ACFs such as in flare2 of Mrk~501 in Figure 
\ref{fig:ccf_mrk501}(b), a sharp peak around zero-lag is clearly seen.
This is most likely due to the ``correlated error'', which can generate 
a zero-lag peak in the DCF or MMD as well, when the compared two time 
series are obtained from a single set of observations.  However, as 
pointed our by \citet{peterson98}, the artificial correlation at 
zero-lag due to correlated errors would drive the measured 
cross-correlation lags to smaller values. Thus, even if correlated 
errors were to affect the data, our results would have been a conservative 
estimate of the significance of a inter-band lag larger than zero. 

\subsection{Uncertainties in the Lags}
As discussed in \citet{edelson01}, there are concerns about measured 
lags that are smaller than the orbital periods in low-Earth orbit satellites.
Periodic interruptions are generated due to the Earth occultation, 
and the orbital period is 5.6 -- 5.8 ks, comparable to the 
measured lags.  However, as can be seen from the histograms of the 
DCF or MMDs, the whole peak (determined from the width of the 
auto-correlation) is shifted, or is asymmetric, showing 
more correlation on one side than the other, and there is no
simple instrumental or sampling reason for this.  We emphasize again 
that the flares we are considering are the ones that have the time 
scales which are quite similar to 
the characteristic time scales which dominate the 
variability.  Another important fact in the case of $ASCA$ data is that 
the time series in different energy bands are all taken by the 
same set of instruments (2 GISs and 2 SISs) so that the observing 
periods are $exactly$ simultaneous.  Accordingly, the data gap in the 
DCF or MMD peak is symmetrical with respect to the zero lag, and thus
there is little chance to measure an artificial lag from time series 
with truly zero-lag.   Thus we conclude that there is little effect 
of the orbital gap on the measured lags.

Another concern about the significance of the lags is 
the effect of the underlying component.  For instance, if a flare 
is located on top of a rising underlying component, 
the peak of the flare would appear to be delayed in time. 
And if this increase is relatively steeper (i.e. as compared to
the amplitude of the flare in its energy band) for the 
soft energy band, an erroneous soft lag will be inferred 
if the offset component is ignored.  From the structure function 
analysis using data sets covering longer time spans, we know that 
there is variability on time scales up to years in blazars 
\citep{kataoka01}, and thus a reliable measurement of inter-band lags 
for a specific flare requires that the offset component is subtracted.

We tested this for the Mrk~501 data, where the flares are clearly 
separated and the offset is rather well defined. This correction 
was not possible for previous data, where the typical observation 
times were $\sim$ a day and the long-term trend couldn't be reliably 
established.  We first estimated the underlying component 
by fitting the light curve on both sides of the flare with 
a linear function for each energy band. We then calculated the lags 
in the same manner as above, after subtracting the underlying component.
The result is shown in Table \ref{tbl:lag_flare}(b), where the estimated 
value of the lag became smaller, but there is still a significant 
lag for flare1 and flare2.  Since the shape of the underlying 
component is not determined unambiguously, we also attempted modeling 
the offset with a second order polynomial and an exponential, but 
the lag remained.  In summary, although there is still possibility 
that the measured lags are artificial due to the underlying component 
which we cannot determine exactly, we conclude that assuming a simple 
linear offset, the lag is real.  A follow up discussion for the 
lags inferred previously from the $ASCA$ data for Mrk~421 \citep{tad96} 
and PKS~2155--304 \citep{kataoka00} is given in Appendix A.

Another important result is that inter-band lags are $not$ universal.  From 
the multiple flares detected in the long look observations, 
we showed that the energy dependence of the lags can differ
from flare to flare.  This supports the indication inferred for 
Mrk~421 \citep{tad00} that in TeV blazars, the intrinsic variability 
time scales including acceleration and 
cooling of the electrons radiating in the X--ray range are comparable, 
and that the balance of these parameters controls the observed lags. 

\section{SUMMARY AND CONCLUSIONS}
We have conducted a unprecedented, uninterrupted long-look 
X--ray observations of the 3 brightest TeV Blazars, namely
Mrk~421, Mrk~501, and PKS~2155--304.  The last two observations 
made use of the opportunity of the continuous long-look observations 
in the last phase (AO-8 program) of $ASCA$.  All sources commonly 
show continuous strong  X--ray flaring, despite the relatively 
faint intensity states in the last two sources.

Structure function analysis shows that all TeV blazars have
a break in the structure function at a 
time scale of $\sim$ a day indicating a characteristic 
time scale, and a very steep slope below the characteristic time 
scale suggesting strong suppression of the more rapid variability.
We carried out simulations in order to verify the sensitivity of the 
observed slope and break time scales to assumed parameters.  In 
particular, we assumed that the time series consists of randomly 
occurring flares (``shots'') superposed on each other and this results in  
the observed variability.  With this, we inferred that the time 
scale of the break in the structure function indicates the minimum 
time scale of the individual shots, and the slope at shorter time 
scales than the break becomes steeper than 1 until it finally reaches 2.
On the other hand, we also detected faster substructures having time scale 
of $\sim 10$ ks, but with smaller amplitude which is consistent 
with those not affecting the structure function.  
The entire variability may also consist of such $\sim 10$ ks 
shots, but if that were the case, in order to match the observations,
we argue the shots should not be fully random
but somewhat correlated with each other.
The near-symmetry of the observed flares 
and the suppression of fast variability
indicates that the characteristic time scale  
is determined by the light crossing time, and thus dominating the 
time scales of the day-scale flares.
In this case, since the acceleration and
cooling times of the electrons are calculated to be
significantly shorter than the light-crossing time,
an inhomogeneous electron distribution (or some other geometrical 
effect) is required in order to account for the non-flat-top flares.  

The energy dependent cross-correlations suggest that
inter-band lags are not universal in TeV blazars.  From 
the multiple flares detected in the long-look
observations, we showed that the energy dependence 
of the lags differs from flare to flare.  This supports the 
inference that the time scales of acceleration and 
cooling of the electrons responsible for the X--ray emission 
are comparable for TeV blazars, and the balance of these parameters
controls the lags. On the other hand, 
we pointed out that some uncertainties remain, 
mainly concerning the nature of the underlying component of 
variability on even longer time scales than considered here.  

\acknowledgements
Support for this work was provided by the Fellowship of
Japan Society for Promotion of Science for Young Scientists,
and NASA through grant number GO6363.01-95A
from the Space Telescope Science Institute, which is operated
by AURA, Inc., under NASA contract NAS~5-26555.  
CT and GM acknowledge support from
Smithsonian Astrophysical Observatory via SAO grant number GO0-1038A.

%% Appendix material should be preceded with a single \appendix command.
%% There should be a \section command for each appendix. Mark appendix
%% subsections with the same markup you use in the main body of the paper.

%% Each Appendix (indicated with \section) will be lettered A, B, C, etc.
%% The equation counter will reset when it encounters the \appendix
%% command and will number appendix equations (A1), (A2), etc.

\appendix

\section{COMMENTS ON THE LAGS PREVIOUSLY MEASURED IN THE 
X--RAY DATA FOR MRK~421 AND PKS~2155--304}
Here we provide a follow-up discussion concerning the significance of the 
the previously published analysis of lags on Mrk~421 and PKS~2155--304
\citep{tad96,kataoka00}.  
In similarity to the flares observed in the long-look observations,
we separated the light curves into 4 energy bands corresponding to 
0.6 -- 1.2 keV, 1.2 -- 1.8 keV, 1.8 -- 3 keV, and 3 -- 10 keV, 
and binned all light curves in 1024 sec bins.  
The DCF and MMD are calculated for the lower three energy bands
against the 3 -- 10 keV band, and the error is 
estimated by the same simulations (see text).
The case of comparing 0.6 -- 1.2 keV and 3 -- 10 keV 
is shown in Figure~\ref{fig:ccf_94}.  
We can see that there is strong correlation for both sources
with $r_{\rm max} > 0.95$,
and that the whole cross-correlation peak is shifted to
the positive side, indicating a soft-lag.
The value of the lag is estimated as the
peak value of a Gaussian fitting the correlation peak.
The derived parameters including the other energy bands
are summarized in Table \ref{tbl:lag_94}.
We also suggest that the low value of correlation ($r_{\rm max}$)
of PKS~2155--304 in \cite{zhang99} pointed out by \cite{edelson01} 
is probably due to the fine binning of the light curve.

As described in the text, there are concerns about the
underlying component, which was assumed to be constant
in previous analysis.
We tested this by assuming a linear underlying component 
with a positive inclination for the soft energy band light curve.
We varied the relative amplitude from 0 to 100 \%, 
subtracted it from the light curve, and calculated the
cross-correlations against the 3 -- 10 keV light curve.
For the 3 -- 10 keV light curve, we subtracted only a constant 
offset for simplicity.  Figure~\ref{fig:offset94} shows the
measured lag between the 0.6 -- 1.2 keV and 3 -- 10 keV light curve
as a function of the assumed percentage of the rise of the 
underlying component for both Mrk~421 and PKS~2155--304.
As expected, it is clearly seen that the measured lag becomes 
smaller as the assumed inclination is higher,
and that a rising component with an increase of $\sim$70 \% for Mrk~421
and $\sim$60 \% for PKS~2155--304 
is needed to cancel out the observed lag.
In other words, if there is an underlying component 
rising with an amplitude of 60 -- 70 \%,
we will obtain the value of lag that we measured,
when in reality no lag exists for the flare itself.
On the other hand, it is likely that the hard energy band would
also rise when the lower energy band rises,
and accordingly this percentage is certainly a minimum value;
thus at least 60 -- 70 \% rise is required to 
cancel out the observed lag. 
An underlying component increasing with such a large amplitude
is not very often observed, but cannot be ruled out.

%% Generally speaking, only the figure captions, and not the figures
%% themselves, are included in electronic manuscript submissions.
%% Use \figcaption to format your figure captions. They should begin on a
%% new page.

\clearpage

%% No more than seven \figcaption commands are allowed per page,
%% so if you have more than seven captions, insert a \clearpage
%% after every seventh one.

%% There must be a \figcaption command for each legend. Key the text of the
%% legend and the optional \label in curly braces. If you wish, you may
%% include the name of the corresponding figure file in square brackets.
%% The label is for identification purposes only. It will not insert the
%% figures themselves into the document.
%% If you want to include your art in the paper, use \plotone.
%% Refer to the on-line documentation for details.

\figcaption[]{X--ray light curve of Mrk~501
binned equally to the satellite orbital 
period during the particular observation.
{\it heavy points:} soft X--ray band (0.6 -- 2~keV); 
{\it light points:} hard X--ray band (2 -- 10~keV).
Each light curve is normalized to its mean count rate,
i.e. 6.0, 3.4 cts/s, respectively. 
A blow-up of light curve during the time region between
the arrow is shown in the box, which demonstrates the
further rapid variability within the day-scale flares.
\label{fig:lc_mrk501}
}

\figcaption[]{Same plot as Figure 1 for PKS~2155--304.
The soft X--ray and hard X--ray light curve is normalized 
to its mean count rate, 7.0, 2.1 cts/s, respectively. 
\label{fig:lc_pks2155}
}
\figcaption[]{Same plot as Figure 1 for Mrk~421.
The soft X--ray and hard X--ray light curve is normalized 
to its mean count rate, 17.6, 5.5 cts/s, respectively. 
\label{fig:lc_mrk421}
}

\figcaption[]{Structure functions for 
(a) Mrk~501, (b) PKS~2155--304, and (c) Mrk~421, 
calculated from the X--ray light curves.
All TeV blazars appear to have structure functions with 
a steep rise at shorter time scales
and a break at a characteristic time scale.
\label{fig:sf}
}

\figcaption[]{The simulated light curves at following conditions;
(a) When all shots have a single time scale; $\tau = 10$ (a-1)
and 100 (a-2).
(b) When all shots have a single time scale $\tau = 50$,
and the average rate of shots per time unit is varied:  
R = 0.1 (b-1), 1 (b-2), 10 (b-3).
(c-1) When each shot is characterized by a random
time scale varying from 10 to 100.  
(d) When the rise and decay time scale differ.
The decay time scale is fixed to $\tau_{\rm d} = 100$,
and the rise time $\tau_{\rm r} = 1$ (d-1), and to 10 (d-2).
\label{fig:sim-lc}
}

\figcaption[]{Structure functions (top panel) and 
their power-law indices (lower panel) for the simulated light curves
shown in Figure~3 under the following conditions:  
(a) When all shots have a single time scale $\tau$.
The crosses show the structure function for $\tau = 10$,
and the diamonds are for $\tau = 100$.
The former is multiplied by a factor of 10;  
the break position indicates the time scale of individual shots.  
(b) When all shots have a single time scale $\tau = 100$,
and the average rate of shots per time unit is varied.  
We can see that the break position does not change as a result 
of varying the rate.
(c) When the shots have a random time scale varying from 10 to 100.
The case with only $\tau = 100$ is shown as crosses,
multiplied by a factor of 10.  
(d) When the rise and decay time scales differ.
The comparison is with the decay time scale fixed to $\tau_{\rm d} = 100$,
and the rise time $\tau_{\rm r} = 1$, 10, and 100.
\label{fig:sim-sf}
}

\figcaption[]{The effect of small amplitude short shots  
on the structure function.
(a) a portion of the simulated light curve  
which consists of single time scale shots ($\tau = 100$), and
(b) with a small and short time scale ($\tau = 10$) shots added on top.
(c) The calculated structure function and its index 
from the light curve with and without the short time scale flares added.
Because the structure function is dominated by larger amplitude
variations, the shorter time scale component will not 
appear in the structure function.
\label{fig:sim-sf2}
}

\figcaption[]{The discrete ACF of the 3 -- 10 keV light curve (top),
and the DCF (middle) and MMD (bottom) between the 
0.6 -- 1.2 keV and 3 -- 10 keV band 
calculated for the entire light curve 
are shown for all three sources:  
(a) Mrk~501, (b) PKS~2155--304, and (c) Mrk~421.
A positive time lag in the DCF and MMD plot 
indicates the 0.6 -- 1.2 keV band delayed from the 3 -- 10 keV band.
\label{fig:ccf_full}
}

\figcaption[]{The discrete ACF of the 3 -- 10 keV light curve (top),
and the DCF (middle) and MMD (bottom) between the 
0.6 -- 1.2 keV and 3 -- 10 keV band
calculated for the individual flares in PKS~2155--304.
A positive time lag in the DCF and MMD plot 
indicates the 0.6 -- 1.2 keV band delayed from the 3 -- 10 keV band.
\label{fig:ccf_mrk501}
}

\figcaption[]{The discrete ACF of the 3 -- 10 keV light curve (top),
and the DCF (middle) and MMD (bottom) between the 
0.6 -- 1.2 keV and 3 -- 10 keV bands 
calculated for the individual flares in Mrk~501.
A positive time lag in the DCF and MMD plot 
indicates the 0.6 -- 1.2 keV band delaying from the 3 -- 10 keV band.
\label{fig:ccf_pks2155}
}

\figcaption[]{The discrete ACF of the 3--10 keV light curve (top),
and the DCF (middle) and MMD (bottom) between the 
0.6 -- 1.2 keV and 3 -- 10 keV band
calculated for the 1994 $ASCA$ data of
(a) Mrk~421 and (b) PKS~2155--304.
A positive time lag in the DCF and MMD plot 
indicates the 0.6 -- 1.2 keV band delayed from the 3 -- 10 keV band.
\label{fig:ccf_94}
}

\figcaption[]{The measured soft lag of the 0.6 -- 1.2 keV band
against the 3 -- 10 keV band as a function of the
assumed amplitude of the underlying offset component
for Mrk~421 (filled circles) and PKS~2155--304 (open circles).
It is shown that a increase with an amplitude of 50 -- 70 \%
is needed to cancel out the obtained lag.
\label{fig:offset94}
}

\clearpage

\begin{deluxetable}{cccccccccccc}
\tabletypesize{\scriptsize}
\tablewidth{0pt}
\tablecaption{Energy dependent cross-correlation results for the full light curve \label{tbl:lag_full} }
\tablehead{
\colhead{} & \multicolumn{3}{c}{Mrk~421}  & \colhead{} &  
\multicolumn{3}{c}{Mrk~501} & \colhead{} &
\multicolumn{3}{c}{PKS~2155--304} \\
\cline{2-4} \cline{6-8} \cline{10-12} \\
\colhead{E\tablenotemark{a}}& 
\colhead{DCF} & \colhead{DCF} & \colhead{MMD} & \colhead{} &
\colhead{DCF} & \colhead{DCF} & \colhead{MMD} & \colhead{} &
\colhead{DCF} & \colhead{DCF} & \colhead{MMD} \\
\colhead{(keV)} &
\colhead{$r_{\rm max}$} & \colhead{peak\tablenotemark{b}} & \colhead{peak\tablenotemark{b}} & \colhead{} &
\colhead{$r_{\rm max}$} & \colhead{peak} & \colhead{peak} & \colhead{} &
\colhead{$r_{\rm max}$} & \colhead{peak} & \colhead{peak} \\
}
\startdata
0.6-1.2 & 0.80 & -1.85$\pm$0.76  & -0.13$\pm$0.52  && 0.78 & -0.06$\pm$0.83  & -2.59$\pm$1.42  && 1.00 & 3.87$\pm$0.98  & 3.62$\pm$0.93  	\\
1.2-1.8 & 0.87 & -0.66$\pm$0.42  & 0.23$\pm$0.26  && 0.87 & 1.02$\pm$0.56  & -0.21$\pm$0.87  && 1.00 & 2.76$\pm$0.80  & 2.13$\pm$0.72  	\\
1.8-3.0 & 0.89 & -0.11$\pm$0.26  & 0.30$\pm$0.15  && 0.90 & 0.68$\pm$0.47  & 0.26$\pm$0.62  && 1.01 & 1.44$\pm$0.64  & 1.37$\pm$0.48  	\\
\enddata
\tablenotetext{a}{All results show the correlation calculated 
against the 3 -- 10 keV light curve}
\tablenotetext{b}{Positive values indicate the soft energy band 
lagging the 3 -- 10 keV light curve in units of kiloseconds.}
\end{deluxetable}

%\clearpage
\begin{deluxetable}{cccccccccccc}
\tablecolumns{12} 
\tabletypesize{\scriptsize}
\tablewidth{0pt}
\tablecaption{Energy dependent cross-correlation results for the individual flares in Mrk~501 and PKS~2155--304 \label{tbl:lag_flare} }
\tablehead{
\colhead{} & \multicolumn{3}{c}{Flare1}  & \colhead{} &  
\multicolumn{3}{c}{Flare2} & \colhead{} &
\multicolumn{3}{c}{Flare3} \\
\cline{2-4} \cline{6-8} \cline{10-12} \\
\colhead{E\tablenotemark{a}}& 
\colhead{DCF} & \colhead{DCF} & \colhead{MMD} & \colhead{} &
\colhead{DCF} & \colhead{DCF} & \colhead{MMD} & \colhead{} &
\colhead{DCF} & \colhead{DCF} & \colhead{MMD} \\
\colhead{(keV)} &
\colhead{$r_{\rm max}$} & \colhead{peak\tablenotemark{b}} & \colhead{peak\tablenotemark{b}} & \colhead{} &
\colhead{$r_{\rm max}$} & \colhead{peak} & \colhead{peak} & \colhead{} &
\colhead{$r_{\rm max}$} & \colhead{peak} & \colhead{peak} \\
}
\startdata
\multicolumn{12}{c}{(a) Mrk~501} \\
\tableline
0.6-1.2  &  0.99 & 2.58$\pm$0.97 & 1.75$\pm$1.25  & &  0.57 & -8.64$\pm$2.00 & -7.36$\pm$2.59  & &  0.61 & 0.52$\pm$1.47 & 3.10$\pm$2.08  \\
1.2-1.8  &  1.00 & 1.98$\pm$0.63 & 1.99$\pm$0.86  & &  0.65 & -5.60$\pm$1.74 & -4.98$\pm$1.97  & &  0.82 & 2.93$\pm$0.92 & 3.58$\pm$1.11  \\
1.8-3.0  &  1.04 & 1.40$\pm$0.52 & 1.66$\pm$0.56  & &  0.74 & -3.27$\pm$1.37 & -3.18$\pm$1.51  & &  0.91 & 2.63$\pm$0.83 & 2.95$\pm$0.88  \\
\cutinhead{(b) Mrk~501 (linear offset subtracted)}
0.6-1.2  &  1.01 & 2.51$\pm$1.00 & 1.95$\pm$1.19  & &  0.59 & -5.85$\pm$1.84 & -3.06$\pm$2.30  & &  0.62 & -1.22$\pm$1.43 & 0.90$\pm$1.91  \\
1.2-1.8  &  1.02 & 1.50$\pm$0.64 & 1.67$\pm$0.81  & &  0.70 & -3.33$\pm$1.58 & -2.04$\pm$1.58  & &  0.85 & 1.97$\pm$0.83 & 2.44$\pm$0.95  \\
1.8-3.0  &  1.06 & 0.82$\pm$0.54 & 1.21$\pm$0.57  & &  0.77 & -1.69$\pm$1.27 & -1.41$\pm$1.16  & &  0.93 & 1.95$\pm$0.75 & 2.16$\pm$0.78  \\
\cutinhead{(c) PKS~2155--304}
0.6-1.2  &  0.72 & 6.06$\pm$2.48 & 6.76$\pm$2.81  & &  0.77 & 9.60$\pm$2.30 & 6.85$\pm$2.44  & &  0.97 & 0.60$\pm$0.69 & 1.35$\pm$0.59  \\
1.2-1.8  &  0.72 & 3.78$\pm$2.32 & 4.72$\pm$2.50  & &  0.80 & 7.95$\pm$2.18 & 4.06$\pm$2.33  & &  0.98 & 0.28$\pm$0.67 & 0.74$\pm$0.58  \\
1.8-3.0  &  0.73 & 3.95$\pm$2.13 & 2.49$\pm$1.84  & &  0.79 & 4.05$\pm$2.08 & 2.54$\pm$1.94  & &  0.97 & -0.20$\pm$0.59 & 0.73$\pm$0.48  \\
\enddata
\tablenotetext{a}{All results show the correlation calculated 
against the 3 -- 10 keV light curve}
\tablenotetext{b}{Positive values indicate the soft energy band 
lagging the 3 -- 10 keV light curve in units of kiloseconds.}
\end{deluxetable}

%\clearpage
\begin{deluxetable}{cccccccc}
\tablewidth{0pt}
\tabletypesize{\scriptsize}
\tablecaption{Energy dependent cross-correlation results for the 1994 $ASCA$ observation \label{tbl:lag_94} }
\tablehead{
\colhead{} & \multicolumn{3}{c}{Mrk~421}  & \colhead{} &  
\multicolumn{3}{c}{PKS~2155--304}  \\
\cline{2-4} \cline{6-8}  \\
\colhead{E\tablenotemark{a}}& 
\colhead{DCF} & \colhead{DCF} & \colhead{MMD} & \colhead{} &
\colhead{DCF} & \colhead{DCF} & \colhead{MMD} \\
\colhead{(keV)} &
\colhead{$r_{\rm max}$} & \colhead{peak\tablenotemark{b}} & \colhead{peak\tablenotemark{b}} & \colhead{} &
\colhead{$r_{\rm max}$} & \colhead{peak} & \colhead{peak} \\
}
\startdata
0.6-1.2  &  0.98 & 5.38$\pm$1.33  & 6.40$\pm$1.28   & &  1.10 & 3.02$\pm$0.72  & 3.37$\pm$0.54   \\
1.2-1.8  &  0.97 & 4.02$\pm$1.03  & 4.68$\pm$1.24   & &  1.11 & 2.73$\pm$0.59  & 2.82$\pm$0.42   \\
1.8-3.0  &  0.97 & 2.62$\pm$0.62  & 2.56$\pm$0.85   & &  1.10 & 1.83$\pm$0.39  & 1.87$\pm$0.31   \\
\enddata
\tablenotetext{a}{All results show the correlation calculated 
against the 3 -- 10 keV light curve}
\tablenotetext{b}{Positive values indicate the soft energy band 
lagging the 3 -- 10 keV light curve in units of kiloseconds.}
\end{deluxetable}

\clearpage
\begin{figure}
\epsscale{0.5}
\plotone{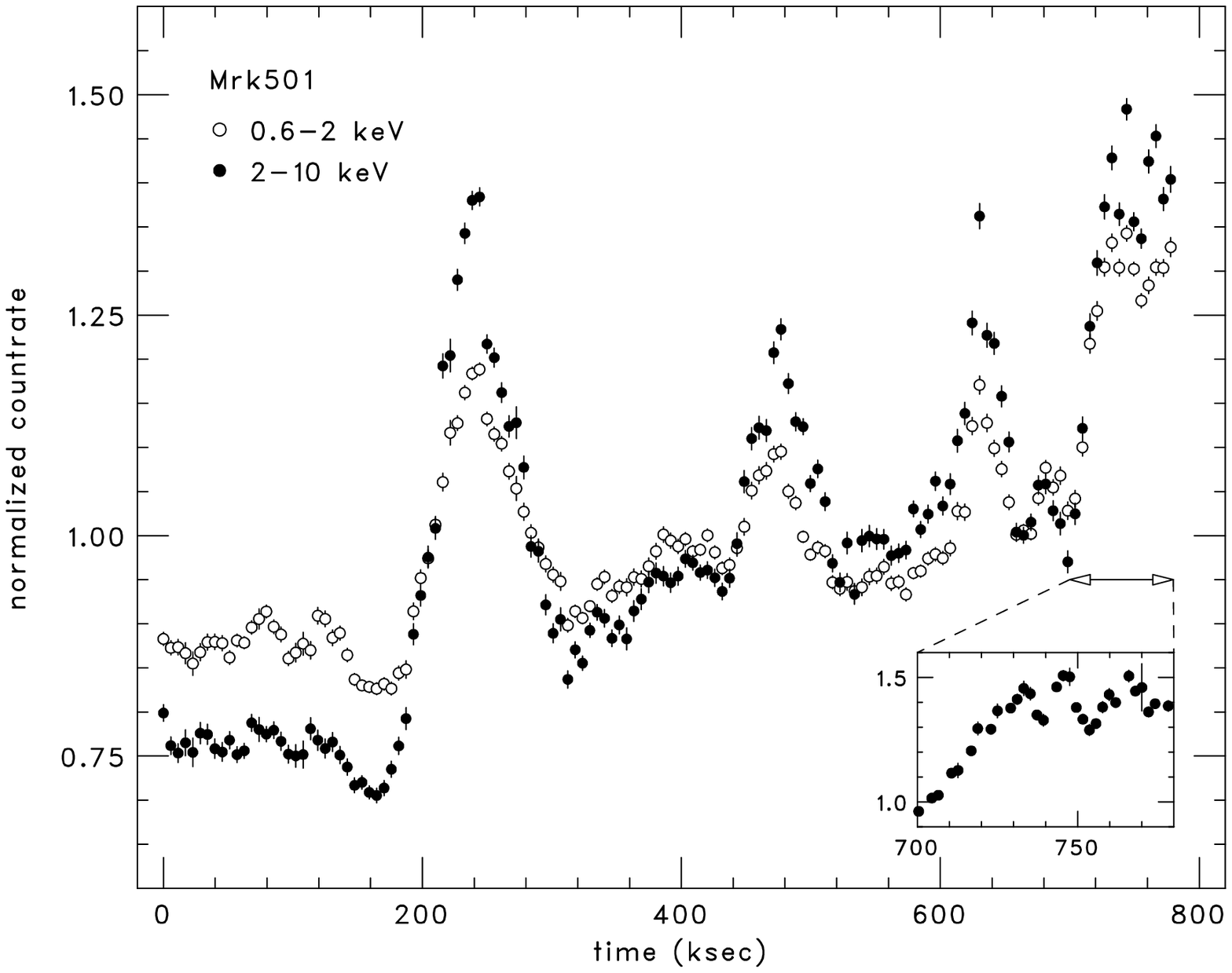}
\end{figure}
%\clearpage
\begin{figure}
\epsscale{0.5}
\plotone{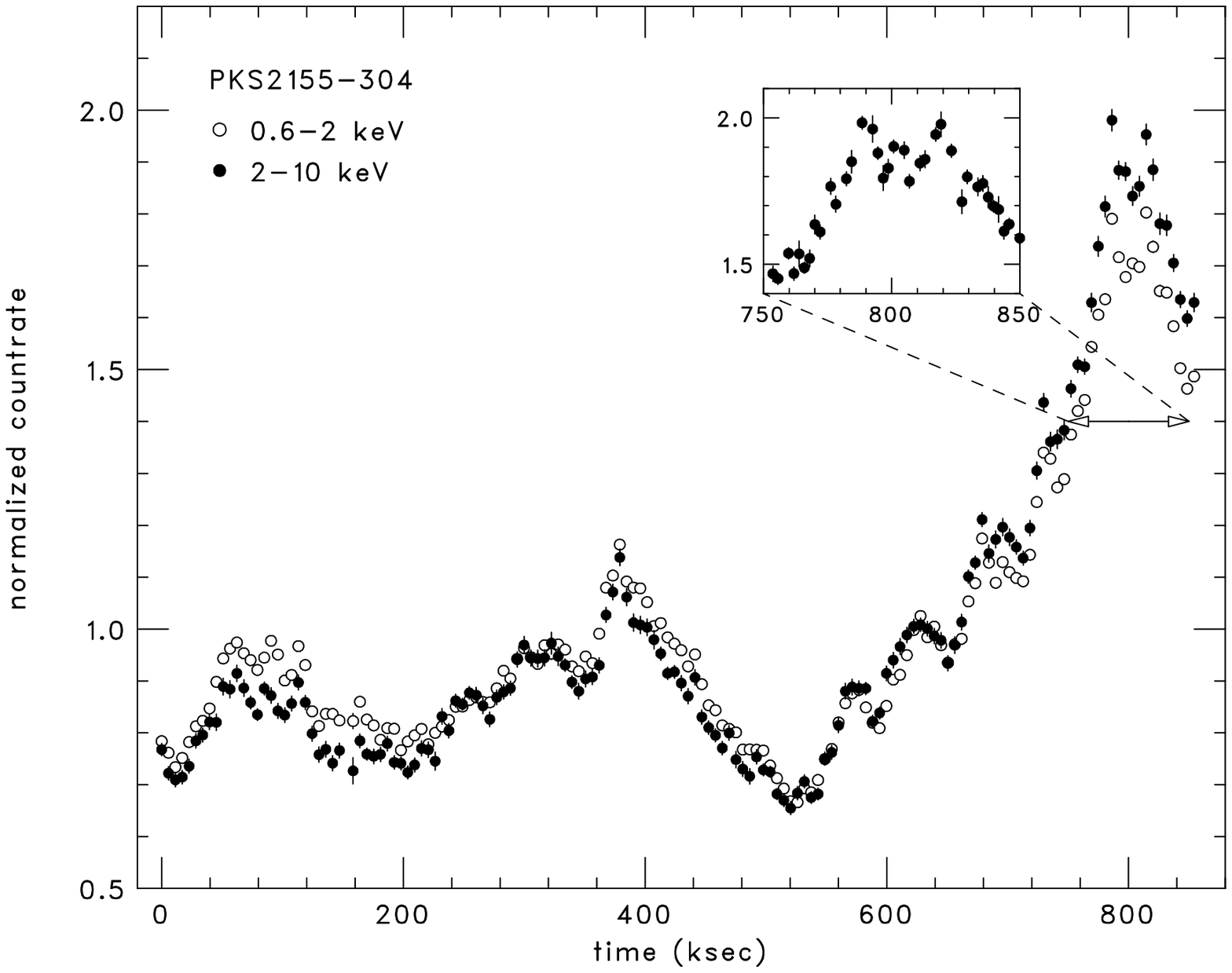}
\end{figure}
%\clearpage
\begin{figure}
\epsscale{0.5}
\plotone{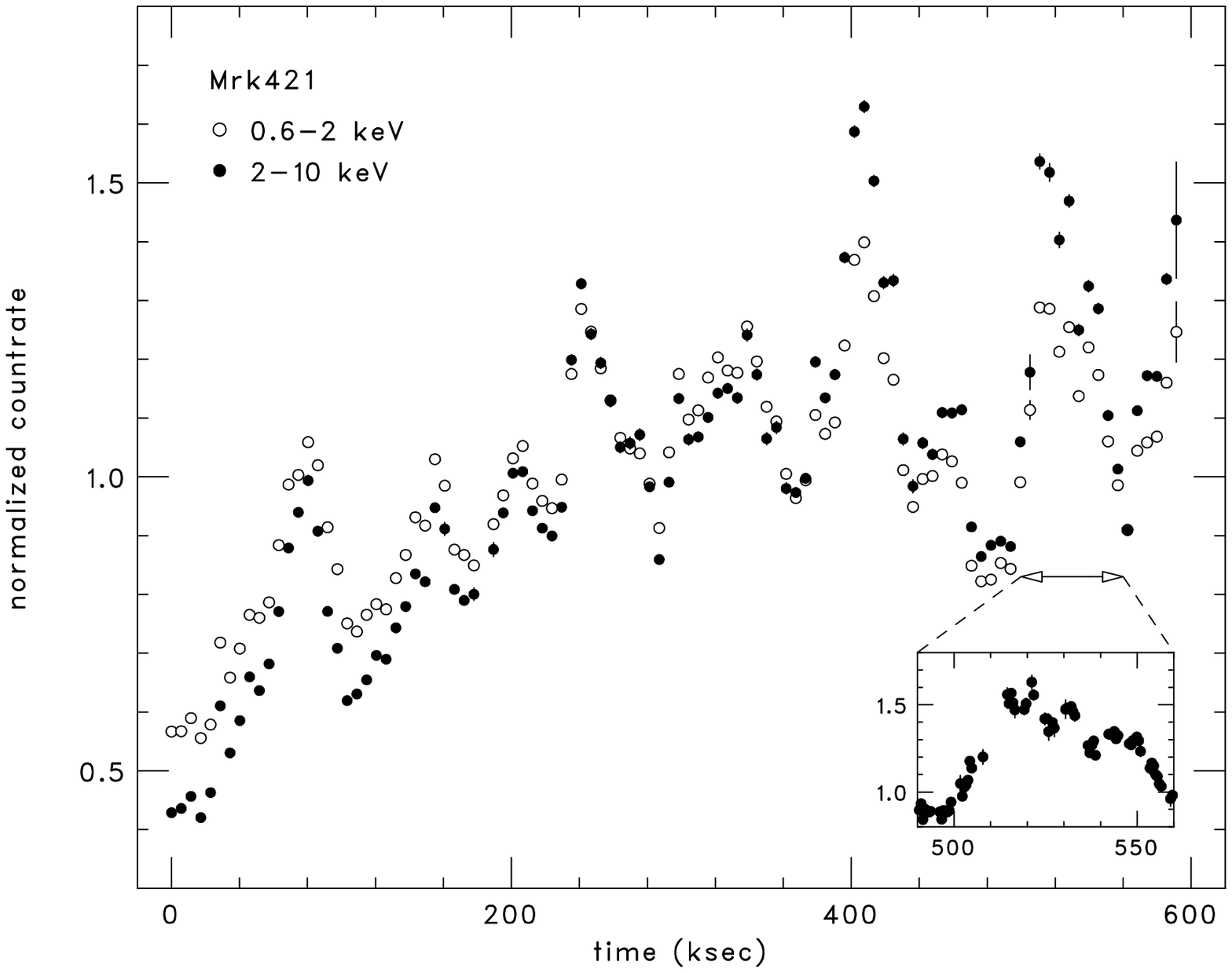}
\end{figure}
%\clearpage
\begin{figure}
\epsscale{0.5}
\plotone{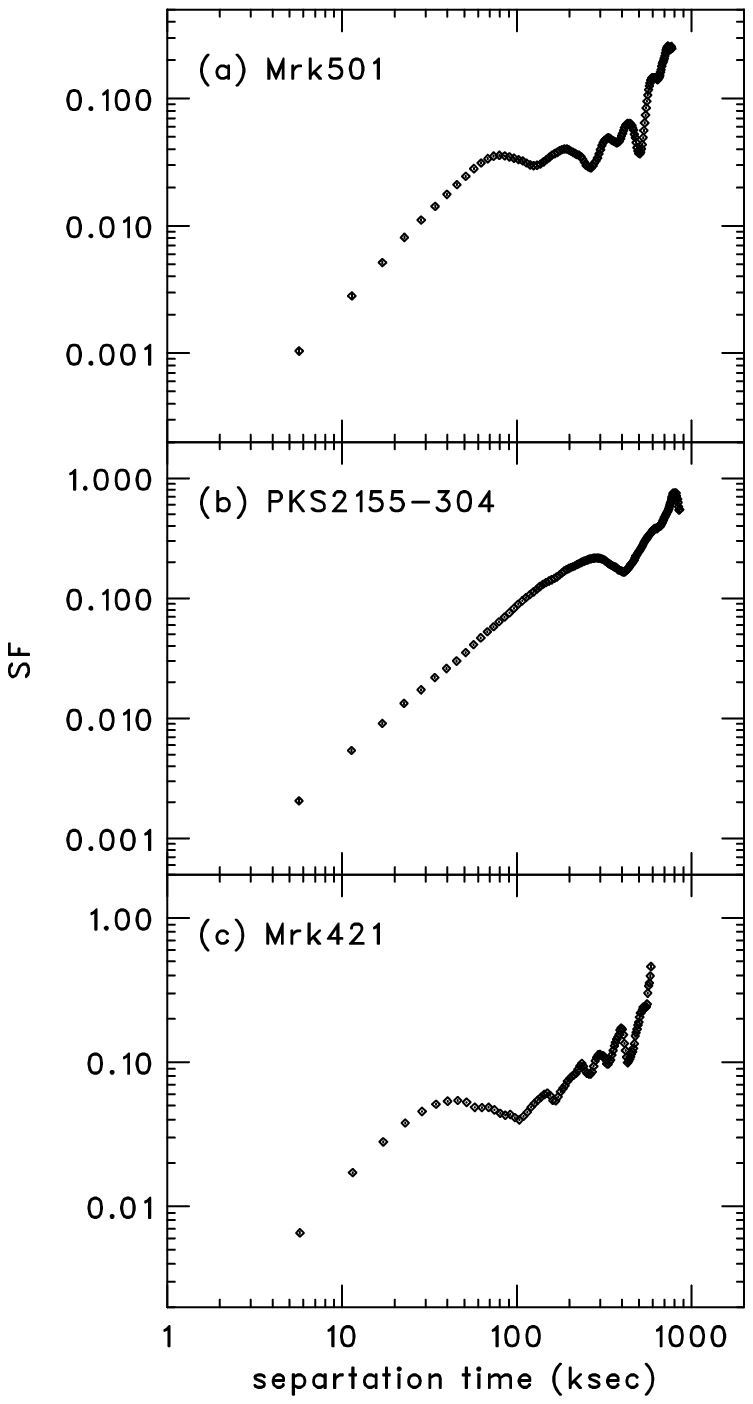}
\end{figure}
%\clearpage
\begin{figure}
\epsscale{0.5}
\plotone{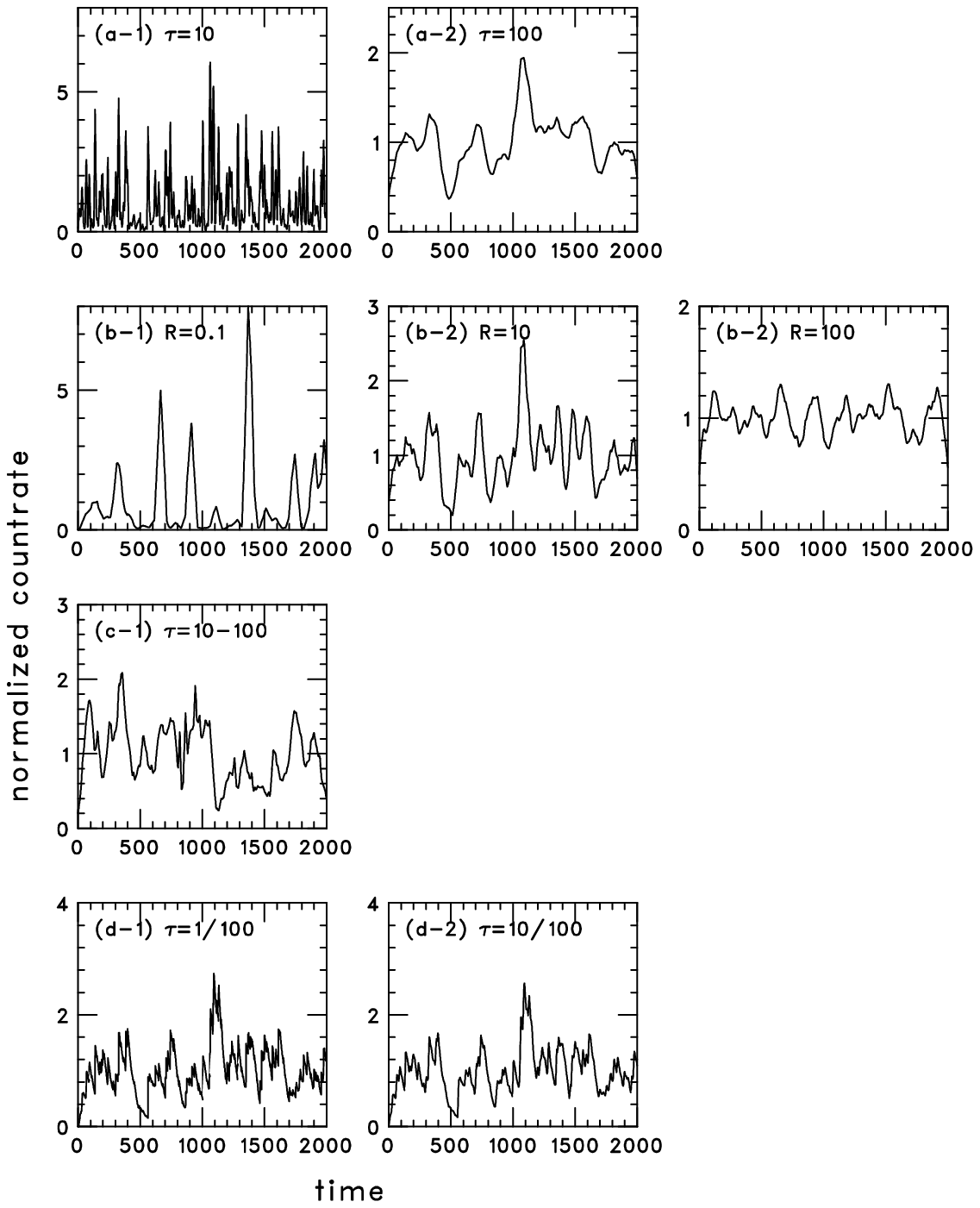}
\end{figure}
%\clearpage
\begin{figure}
\epsscale{0.5}
\plotone{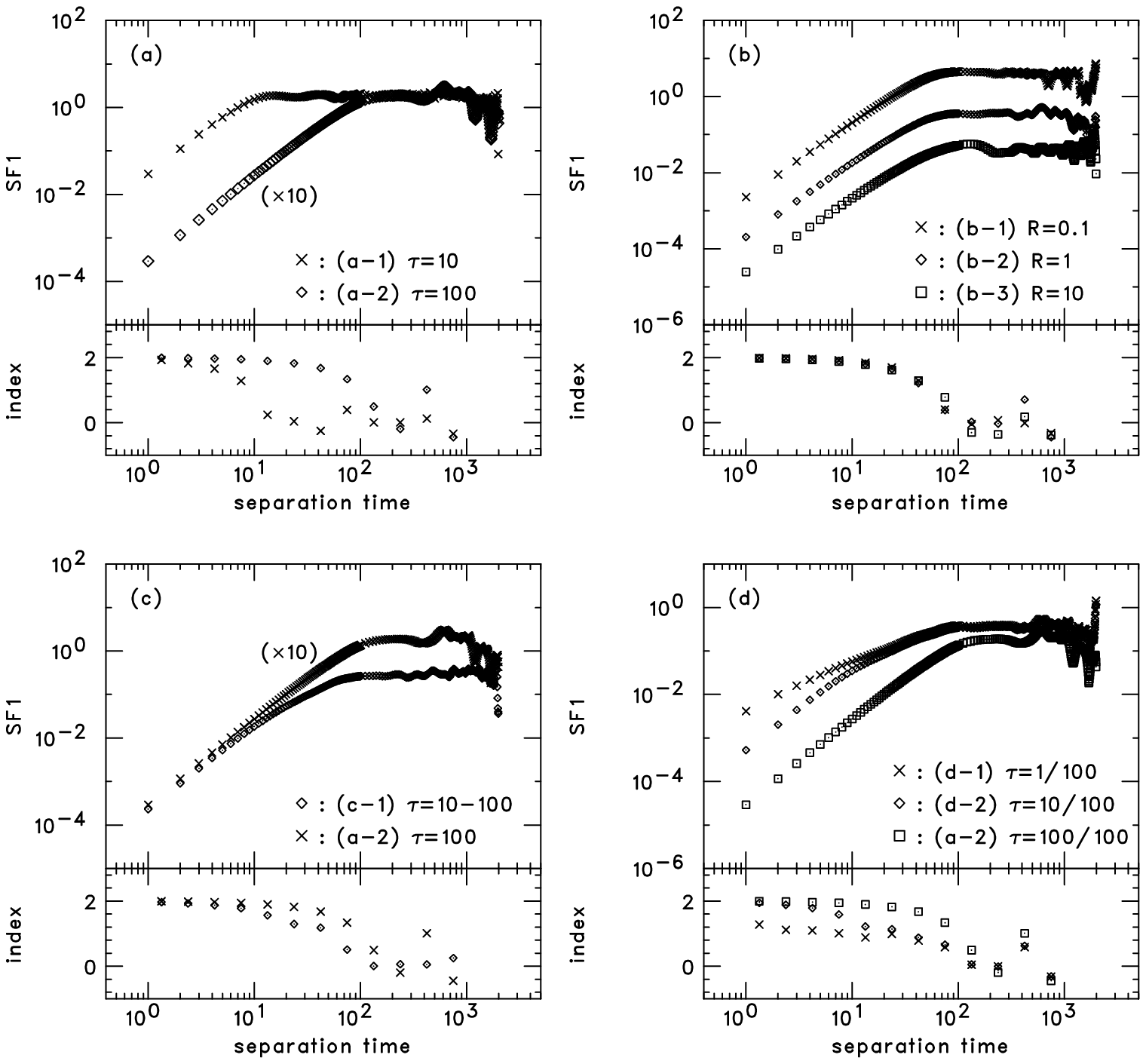}
\end{figure}
%\clearpage
\begin{figure}
\epsscale{0.5}
\plotone{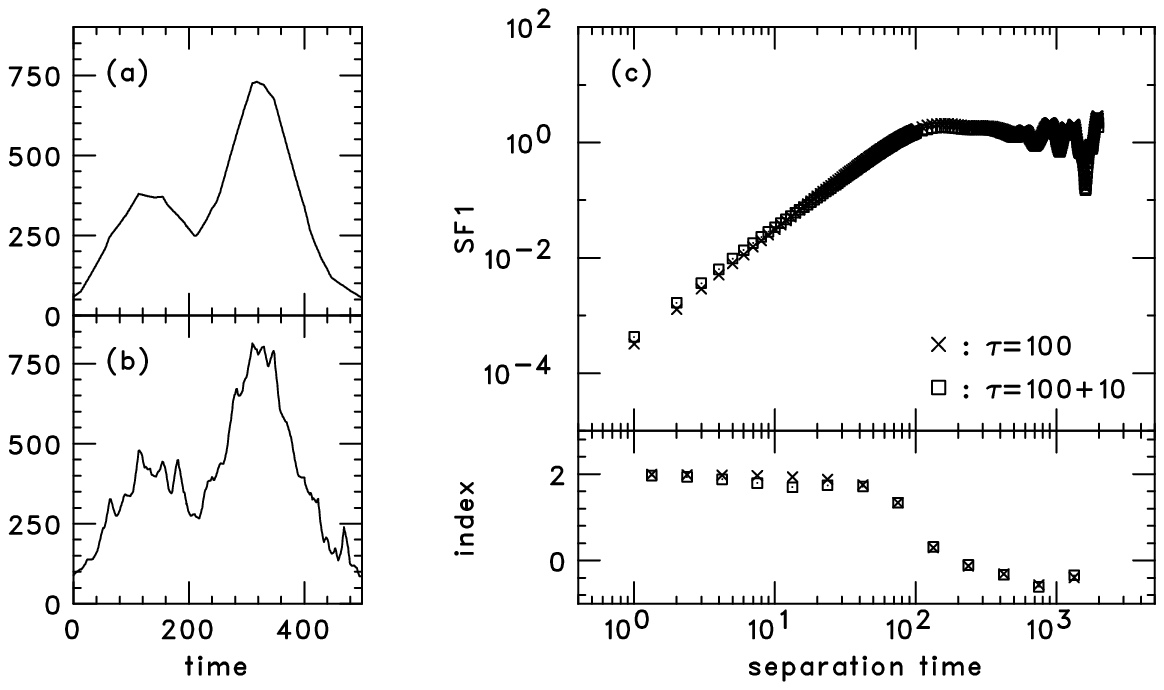}
\end{figure}
%\clearpage
\begin{figure}
\epsscale{0.5}
\plotone{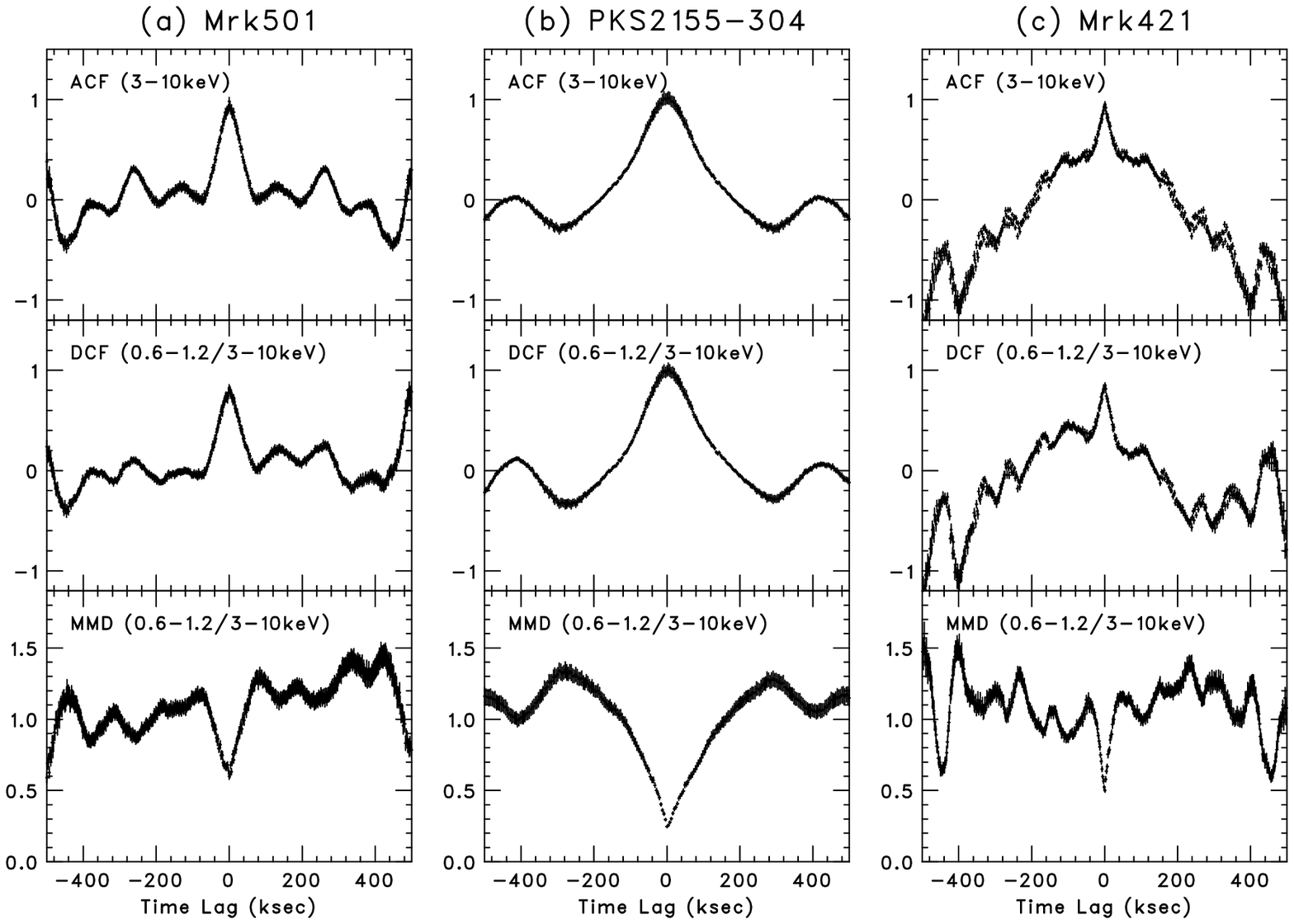}
\end{figure}
%\clearpage
\begin{figure}
\epsscale{0.5}
\plotone{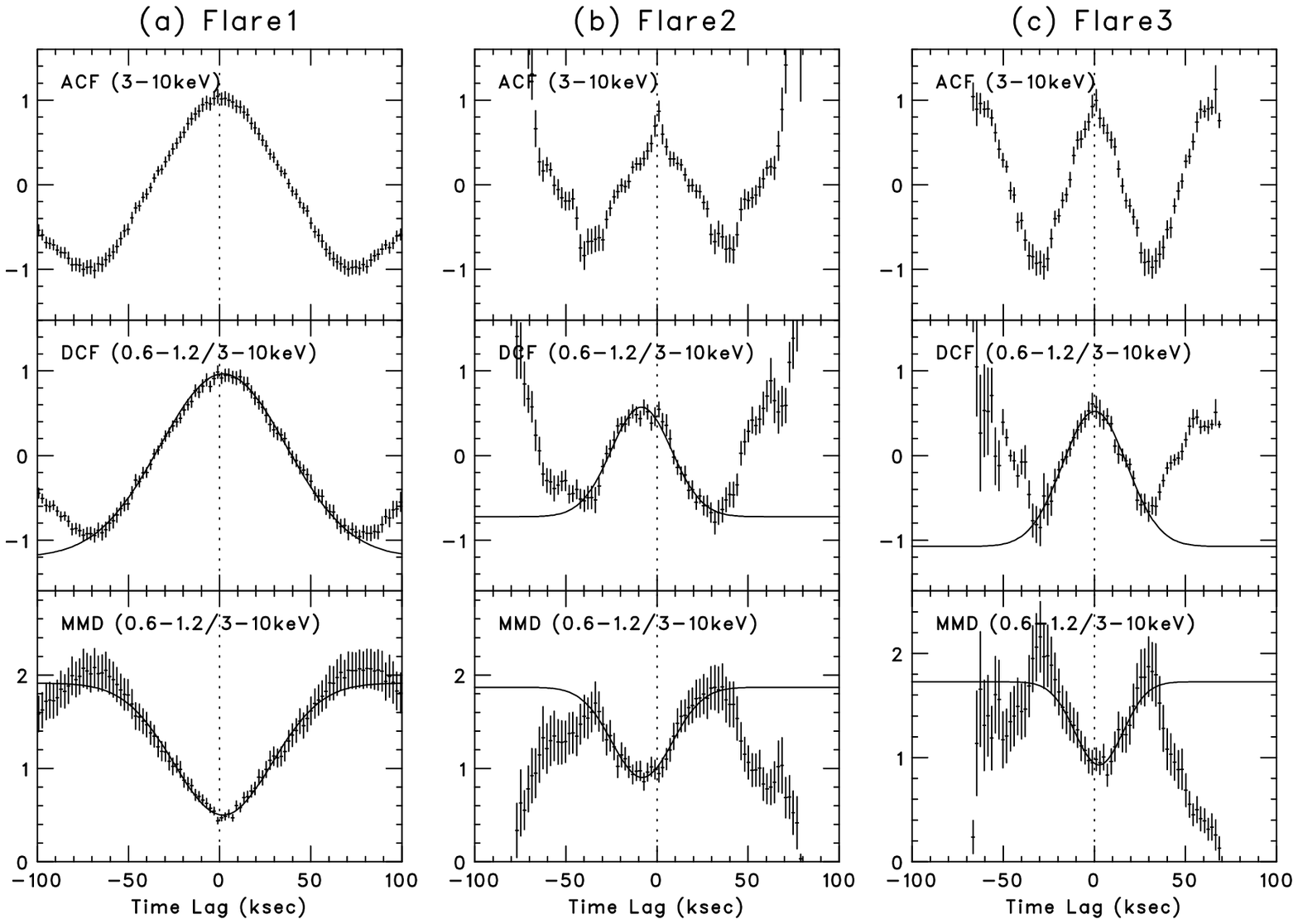}
\end{figure}
%\clearpage
\begin{figure}
\epsscale{0.5}
\plotone{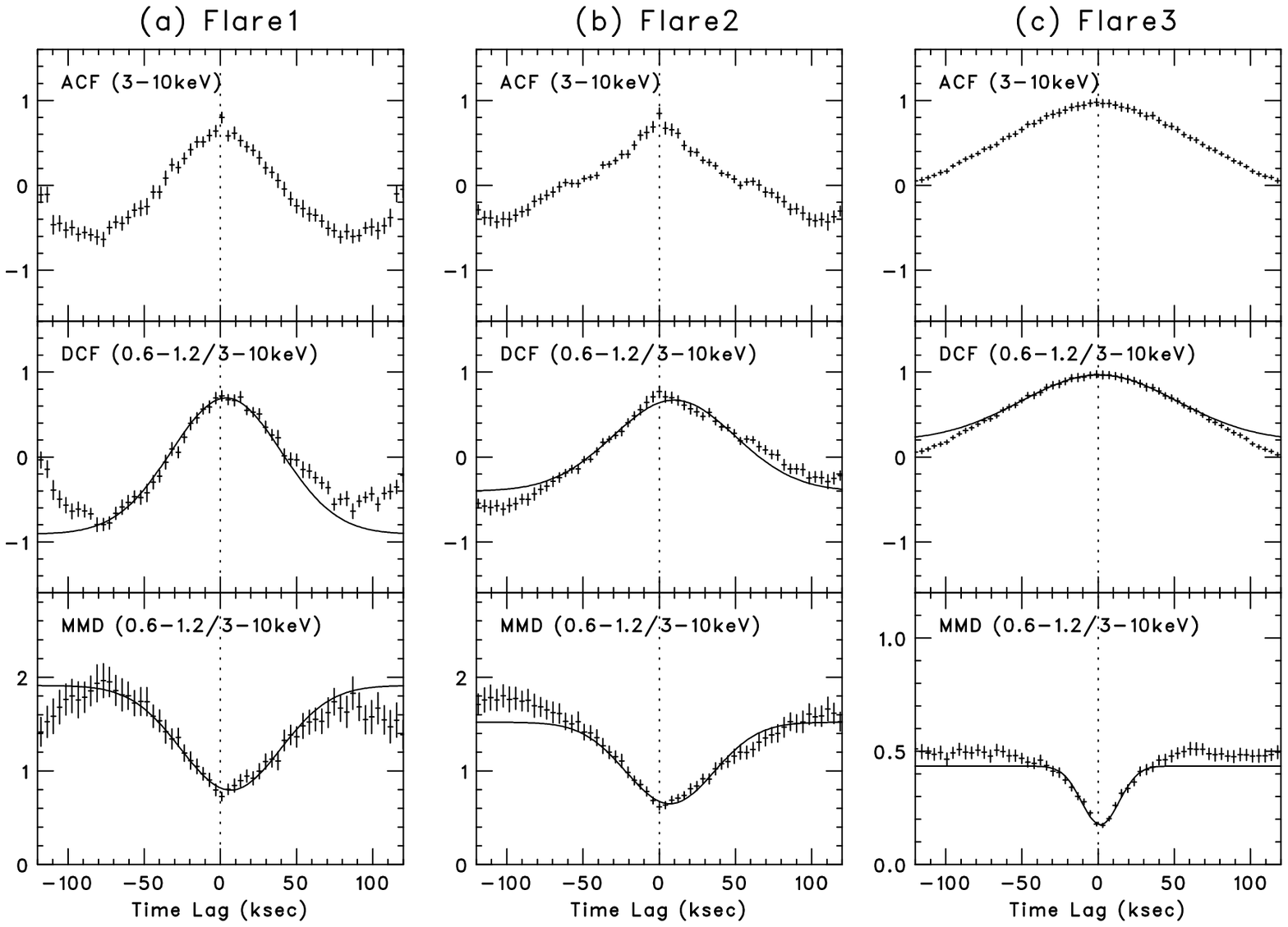}
\end{figure}
%\clearpage
\begin{figure}
\epsscale{0.5}
\plotone{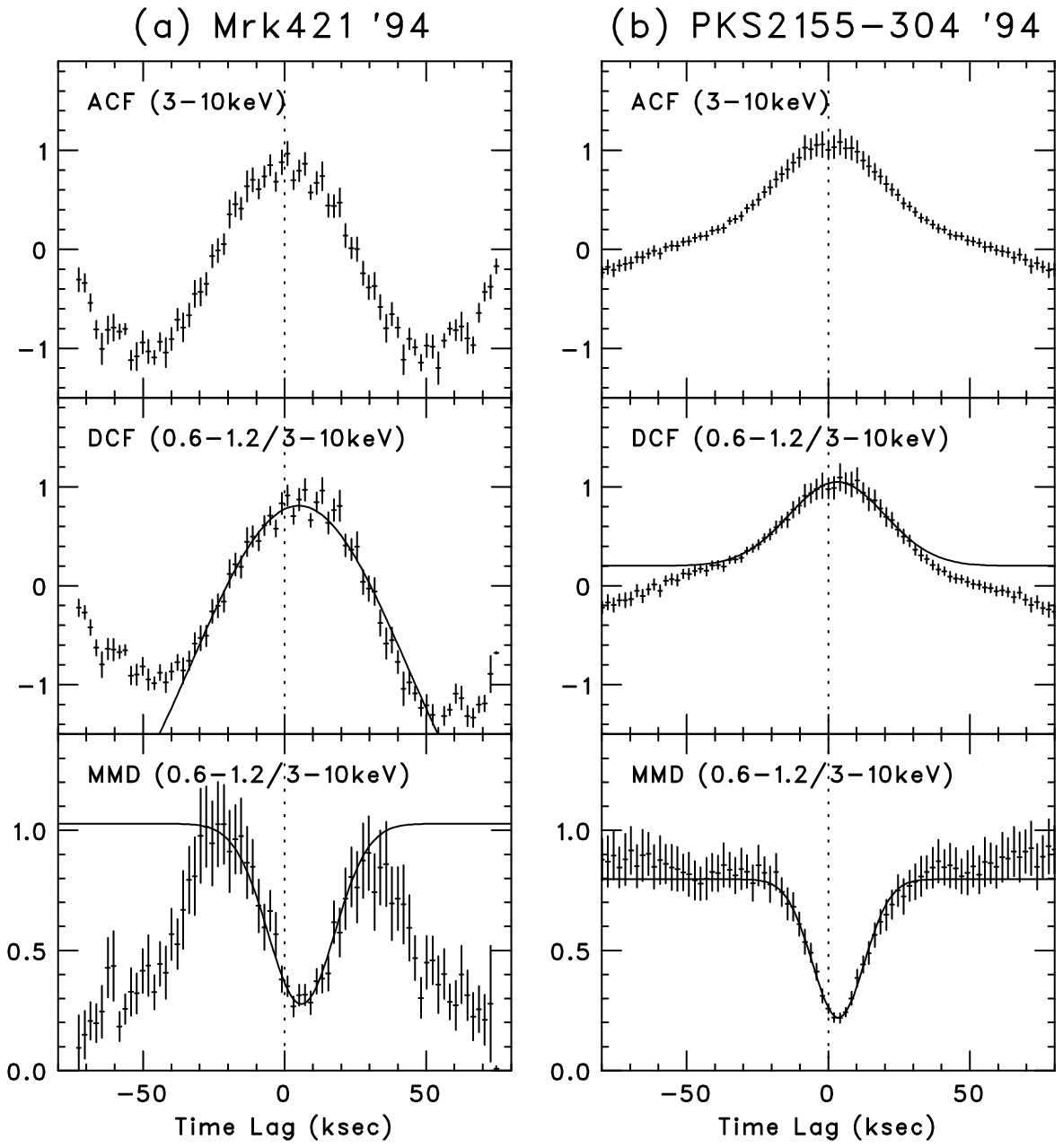}
\end{figure}
%\clearpage
\begin{figure}
\epsscale{0.5}
\plotone{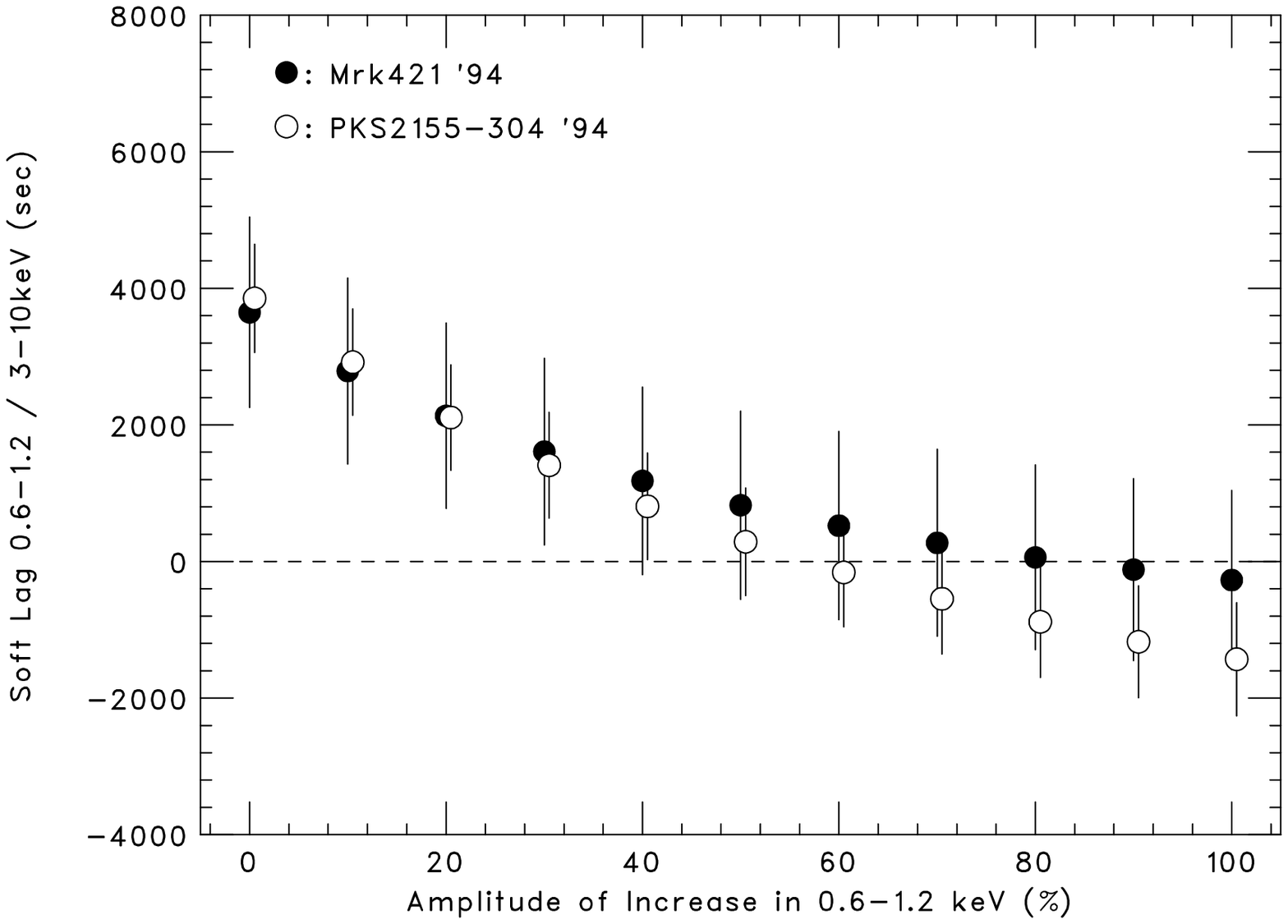}
\end{figure}

\end{document}